\documentclass[%
reprint,
superscriptaddress,
prb,
]{revtex4-2}
\usepackage[english]{babel}
\usepackage{graphicx}
\usepackage{dcolumn}
\usepackage[bottom]{footmisc}
\usepackage{bm}
\usepackage{color}
\usepackage{ulem}
\usepackage{hyperref}

\begin{document}


\title{Complex interplay of magnetic ordering and spin-lattice coupling in orthochromite Nd$_{0.5}$Dy$_{0.5}$CrO$_{3}$}

\author{M. Anas}
\affiliation{Department of Physics, I.I.T. Roorkee-247667, India.}
\author{Padmanabhan Balasubramanian}
 \affiliation{Graphic Era University, Dehradun-248002, India}
\author{K. Vikram}
 \affiliation{Graphic Era University, Dehradun-248002, India}
\author{Ankita Singh}
\altaffiliation{Present address: Department of Condensed Matter Physics and Materials Science, Tata Institute of Fundamental Research, Homi Bhabha Road, Colaba, Mumbai-400005, India.}
\affiliation{Department of Physics, I.I.T. Roorkee-247667, India.}

\author{\mbox{C. M. N. Kumar}}
\affiliation{Institute of Solid State Physics, Vienna University of Technology, Wiedner Hauptstraße 810, 1040 Vienna, Austria}
 \affiliation{AGH University of Science and Technology, Faculty of Physics and Applied Computer Science, 30-059 Karak\'ow, Poland}
\author{Andreas Hoser} 
\affiliation{Helmholtz-Zentrum Berlin, Germany}
\author{Dariusz Rusinek}
\affiliation{National Center for Nuclear Research, Otwock-Swierk, Poland}
\author{A.K. Sinha}
\altaffiliation{Present address: Department of Physics, School of Engineering, UPES, Dehradun-248007, India.}
\affiliation{Indus-2, Raja Ramanna Centre for Advanced Technology, Indore, India}
\author{V. Srihari} 
\affiliation{Indus-2, Raja Ramanna Centre for Advanced Technology, Indore, India}
\author{Ranjan K. Singh}
\affiliation{Department of Physics, Banaras Hindu University, Uttar Pradesh, India}
\author{Rinku Kumar}
\affiliation{Institute Instrumentation Centre, I.I.T. Roorkee-247667, India}
\author{Mukul Gupta}
\affiliation{UGC-DAE Consortium for Scientific Research, University Campus, Khandwa Road, Indore-452001, India}
\author{T. Maitra}
\affiliation{Department of Physics, I.I.T. Roorkee-247667, India.}
\author{V. K. Malik}
\email{vivek.malik@ph.iitr.ac.in }
\affiliation{Department of Physics, I.I.T. Roorkee-247667, India.}





\date{\today}

\begin{abstract}
The mixed rare-earth orthochromite Nd$_{0.5}$Dy$_{0.5}$CrO$_{3}$ has a N\'eel temperature ($T_\mathrm{{N}}$) of ${\sim}$ 175\,K, resulting in the G-type antiferromagnetic ordering of Cr$^{3+}$ spins. The inverse susceptibility shows a deviation from Curie-Weiss law at 230\,K, with a large effective paramagnetic moment of 8.8\,${\mu}_{\mathrm{B}}$. The ZFC-FC magnetization bifurcate just above $T_\mathrm{{N}}$ and show a distinct signature of spin reorientation near 60\,K.
Neutron diffraction show that below $T_\mathrm{{N}}$, the Cr$^{3+}$ spins align in ${\Gamma}_{2}$ representation as ($F_{x}$, $G_{z}$). Below 60\,K, due to spin reorientation, the magnetic structure is in  ${\Gamma}_{1}$ ($G_{y}$) configuration. The neutron diffraction does not show any signature of rare-earth ordering even at 1.5\,K. First principles density functional theory calculations within GGA+U and GGA+U+SO approximations reveal
that the G-type antiferromagnetic order is the ground state magnetic structure of Cr sublattice and the spin-reorientation of Cr$^{3+}$ spins can happen in the absence of 3d-4f interactions unlike in the case of orthoferrites.    
The specific heat shows a `${\lambda}$' anomaly at $T_\mathrm{{N}}$, while at low temperature two distinct Schottky anomalies are observed; a Schottky peak at 2\,K and an additional step-like feature above 10\,K. 
Above $T_\mathrm{{N}}$, the magnetic transition is preceded by structural anomalies as seen in our x-ray diffraction and Raman measurements.
The deviation of structural parameters near N\'eel temperature is smaller.
The phonon frequencies show deviation from the standard anharmonic behaviour: first near 250\,K, due to magneto-volume effects while the second deviation occurs near 200\,K due to spin-phonon coupling.
\end{abstract}

\maketitle
\section{\label{sec:intro}Introduction}
The magnetic properties of orthochromites $R$CrO$_{3}$ (R: rare-earth) have been extensively studied in the past fifty years\cite{Hornreich1978,VanLaar1971,McDannald2016,McDannald2015,McDannald2013}. The magnetic ordering is predominantly G-type antiferromagnetism\cite{Hornreich1978,VanLaar1971,Bertaut1968,McDannald2016}, along with weak ferromagnetism (FM) due to the Dzyaloshinskii-Moriya interaction between the Cr$^{3+}$ spins\cite{McDannald2016,McDannald2015}. 
With decreasing temperature, similar to orthoferrites ($R$FeO$_{3}$), they show the interesting property of spin reorientation, wherein the easy axis of the Cr$^{3+}$ spins rotate from one crystallographic axis to another\cite{yamaguchi1974,Hornreich1978}. The spin reorientation can be induced by external magnetic field or laser source, which finds commercial applications in spin switching devices\cite{levy1971,Kimel2004,Cao2014,Zhang2016}. 
Additional interesting properties of orthochromites include magnetocaloric effect (useful in refrigeration) and the magnetoelectric effect, which is useful in spintronic and other multifunctional devices that can be controlled by electric and magnetic fields\cite{McDannald2013,Yin2016,Kumar2017,Yin2014,Li2019}. 
\newline
Structurally, the orthoferrites and orthochromites crystallize in the orthorhombic $Pbnm$ space group.
Compared to the orthoferrites, which have a N\'eel temperature (\,T$_{\mathrm{N1}}$) between  620\,K and 740\,K (depending on $R$)\cite{Eibschutz1967,WHITE1969,Tsymbal2007}, the orthochromites show paramagnetic (PM)-antiferromagnetic (AFM) ordering of Cr$^{3+}$ sublattice between 290\,K and 120\,K\cite{Hornreich1978}. The Cr$^{3+}$ ions have been observed to order near 220\,K and 146\,K in NdCrO$_3$ and DyCrO$_3$ respectively\cite{Du2010,McDannald2013}.
Similar to Fe$^{3+}$ spins, the magnetic structure of Cr$^{3+}$ spins can be described in Betraut notation as ${\Gamma}_{1}$ ($A_{x}$, $G_{y}$, $C_{z}$), ${\Gamma}_{2}$ ($F_{x}$, $C_{y}$, $G_{z}$), ${\Gamma}_{3}$ ($C_{x}$, $F_{y}$, $A_{z}$) and ${\Gamma}_{4}$ ($G_{x}$, $A_{y}$, $F_{z}$)\cite{bertaut1963magnetism}.
Unlike the orthoferrites, which order in ${\Gamma}_{4}$ below \,T$_{N1}$ \cite{WHITE1969}, the orthochromites can order in ${\Gamma}_{2}$ or ${\Gamma}_{4}$ magnetic structures below the N\'eel temperature\cite{Hornreich1978}. 
Interestingly, the spin reorientation behaviour is also different from that in orthoferrites.
For instance, NdCrO$_{3}$ shows a ${\Gamma}_{2}$${\rightarrow}$${\Gamma}_{1}$ abrupt reorientation near 35\,K while DyCrO$_{3}$ does not show any spin reorientation and remains in the ${\Gamma}_{2}$ structure\cite{Bartolome2000,Hornreich1978,McDannald2016}.  This is contrary to that in orthoferrites which show the canonical ${\Gamma}_{4}$${\rightarrow}$${\Gamma}_{2}$ reorientation\cite{WHITE1969,Yuan2011}, while DyFeO$_{3}$ shows ${\Gamma}_{4}$${\rightarrow}$${\Gamma}_{1}$ reorientation\cite{WHITE1969}.
This suggests that the influence of rare-earth in orthochromites is different as compared to that in orthoferrites.
\newline
Additionally, the rare-earth ordering in $R$CrO$_{3}$ varies across the $R$ series.
Below 10\,K, the rare-earths order independently (for Dy, Tb) or due to polarization of Cr$^{3+}$ spins\cite{Hornreich1978,VanLaar1971}. The temperature at which independent rare-earth ordering occurs can be considered as the second N\'eel temperature ($T_\mathrm{{N2}}$)\cite{Yin2016}.
 In DyCrO$_{3}$, the Dy$^{3+}$ moments order independently as ${\Gamma}_{5}$($g_{x}^R$,$a_{y}^{R}$)\cite{Krynetskii1997}, while the Nd$^{3+}$ moments in NdCrO$_3$ order as $c_{z}^{R}$, which is symmetry compatible with ${\Gamma}_{1}$ structure\cite{Shamir1981}.
In NdCrO$_{3}$, the Nd$^{3+}$-Nd$^{3+}$ interactions are weaker than the Nd$^{3+}$-Cr$^{3+}$ interactions. Therefore, the collective independent ordering of Nd$^{3+}$ moments does not occur\cite{Hornreich1975,Bartolome2000}. 
\newline
Important properties observed in orthoferrites and orthochromites are the magneto-electricity and ferroelectric (FE) polarization\cite{Rajeswaran2012,saha2014}. 
In DyFeO$_{3}$, the breaking of inversion symmetry takes place due to ${\Gamma}_{25}$ magnetic structure, that has non-zero magnetoelectric tensor. In addition to this, the exchange striction mechanism between Dy$^{3+}$ and Fe$^{3+}$ moments has been suggested to be responsible for the origin of ferroelectricity in DyFeO$_{3}$, below its ordering temperature (4\,K)\cite{Tokunaga2008}.
Similar mechanism should be valid for the orthochromites. Thus, the magnetoelectric effect in orthochromites is associated with the rare-earth ordering which occurs below $T_\mathrm{{N2}}$ in DyCrO$_{3}$ and TbCrO$_{3}$\cite{Yin2016}.  
However, this does not clearly elucidate the possibility of ferroelectricity in the system.
%
Experimentally, ferroelectricity has been observed in orthochromites at temperatures below as well as above $T_\mathrm{{N1}}$\cite{SrinuBhadram2013,Indra_2016,Ghosh2014,Sharma2014}. The nature of ferroelectricity is believed to be controversial. In ErCrO$_{3}$, the electric polarization is observed below $T_\mathrm{{N1}}$\cite{Su2012}.
In NdCrO$_{3}$, the ferroelectric phase develops at 88\,K, which is well below the N\'eel temperature\cite{Indra_2016}, while in the case of SmCrO$_{3}$, the ferroelectric phase develops in the paramagnetic region itself\cite{Ghosh2014}. Interestingly, in DyCrO$_{3}$, the FE polarization is not reported, but a paraelectric phase is observed below the antiferromagnetic ordering temperature\cite{Yin2018}.
The development of ferroelectric polarization is also accompanied with the breaking of inversion symmetry of the unit cell, which is a requisite condition\cite{Barone2011,Yin2018}. 
The structural distortion associated with FE polarization is believed to be the result of an off-centric displacement of the Cr$^{3+}$ ions within the oxygen octahedra\cite{Yin2018}.
In addition to high-resolution x-ray diffraction using synchrotron radiation, the local structural changes can also be probed using Raman spectroscopy. Deviation from anharmonicity, especially close to N\'eel temperature, suggests a strong spin-phonon coupling\cite{SrinuBhadram2013}.
\newline
Due to the different magnetic and structural properties shown by NdCrO$_{3}$ and DyCrO$_{3}$, it would be interesting to observe the magnetic structure and the resultant spin-phonon coupling in doped compound Nd$_{0.5}$Dy$_{0.5}$CrO$_{3}$. Thus, the present studies are based on the doped orthochromite Nd$_{0.5}$Dy$_{0.5}$CrO$_{3}$, wherein we explore the magnetic properties viz. spin reorientation, rare-earth ordering and the lattice distortion occuring below 300\,K, deviation  of the phonon modes from anharmonic behaviour observed in the Raman spectra is also studied. 
We have explored the bulk magnetic behaviour by DC magnetization and specific heat measurements. The temperature dependent magnetic structure has been studied using neutron diffraction. The possibility of structural distortion was explored using temperature-dependent synchrotron XRD studies. The behaviour of various phonon modes associated with the R$-$O and Cr$-$O vibrations have been studied using temperature-dependent Raman spectroscopy.

\section{\label{sec:method}Methodology}
\subsection{\label{sec:exp}Experimental}
The polycrystalline samples of Nd$_{0.5}$Dy$_{0.5}$CrO$_{3}$ (NDCO) were synthesised by conventional solid-state reaction method using stoichiometric proportions of highly pure (99.99${\%}$) pre-dried precursors of Nd$_{2}$O$_{3}$, Dy$_{2}$O$_{3}$ and Cr$_{2}$O$_{3}$. The powdered constituents were mixed thoroughly and heated in a sequential manner at 1000\,$^{\circ}$C, 1200\,$^{\circ}$C, 1300\,$^{\circ}$C and finally at 1350\,$^{\circ}$C for 24 hrs. 

The structural phase of the sample was  identified using a Rigaku MiniFlex benchtop x-ray diffractometer employing Cu $K_{\alpha}$ radiation.  
Temperature-dependent x-ray diffraction (XRD) measurements were performed at the synchrotron radiation facility Indus-2, India. The powder diffraction in the temperature range 5\,K to 300\,K was carried out at the Angle dispersive X-ray Diffraction (ADXRD) beamline (BL-12), while the high temperature experiments were carried out at BL-11, in the range 300\,K to 500\,K.
 Rietveld refinement of the diffraction patterns were performed using the FULLPROF package\cite{rodriguez1990fullprof}.

DC magnetization was measured using the SQUID magnetometer of magnetic property measurement system (MPMS-XL) and the vibrating sample magnetometer (VSM) module of physical property measurement system (PPMS-Dynacool) manufactured by Quantum Design Inc. 
Zero field cooled (ZFC) and field cooled (FC) measurements from 300\,K to 2\,K in magnetic fields of 50\,Oe, 100\,Oe and 1000\,Oe were carried out to identify the different magnetic transitions. Field variation of magnetization was measured at various temperatures between 300\,K and 2\,K. 

Heat capacity measurements in the temperature range 2-300\,K were carried out using a QD-PPMS (Physical Property measurement system by Quantum Design). Additional heat capacity measurements in the milli-Kelvin range were carried out using the QD-PPMS with ${}^{3}$He option in external magnetic fields of 0, 2, and 5\,T.
Powder Neutron diffraction studies at zero magnetic field were carried out at various temperatures in the range 1.5-300\,K to identify the crystal, magnetic structure and their variations as a function of temperature. The studies were carried out at focusing powder diffractometer E6 at BER-II reactor in Helmholtz-Zentrum Berlin, Germany ($\lambda$ = 2.40\,$\mathrm{\AA}$).  The diffraction data were analyzed using FullProf \cite{rodriguez1990fullprof} suite of programs using the Rietveld method\cite{rietveld1969profile}. The magnetic structure was determined using the irreducible representations from BasIreps \cite{Hovestreydt:wi0099} and refined using FullProf. 
\newline
Raman spectra of NDCO were recorded in the region 50 to 800 cm$^{-1}$ at different temperatures varying from 300\,K to 80\,K in single channel mode on a micro-Raman setup (Model RM1000) procured from Renishaw, equipped with a 2400\,lines/mm grating and a Peltier cooled CCD. The
532\,nm line of He-Ne laser was used as an excitation source. The resolution of the spectrometer was nearly 1\,cm$^{-1}$. A microscope from Olympus (Model MX50 A/T) attached with the spectrometer focuses the incident laser light
on the sample and collects the back-scattered Raman signal. The sample was kept in a quartz sample holder, which was put in the computer-controlled high/low temperature cell, THM 600 temperature controlled stage from Linkam Scientific Instruments.  The individual peaks were fitted to Lorentzian function.
\subsection{\label{sec:dft}Theoretical}
For understanding of the electronic structure and low temperature magnetic behaviour of NDCO, we have performed density functional theory (DFT) calculations as implemented in the Vienna Ab-initio Simulation Package (VASP), which uses the projector augmented wave (PAW) method\cite{kresse1996efficient}. The calculations were performed using Perdew-Burke-Ernzerhof (PBE) exchange-correlation functional within generalized gradient approximation (GGA)\cite{perdew1996} and GGA+$U$\cite{anisimov} approximation. A cut-off energy of 500\,eV was used in the expansion of the plane waves. A 6$\times$6$\times$6 Monkhorst-Pack $k$-mesh centered at ${\Gamma}$ point in Brillouin zone is used for performing the Brillouin zone integrations.
The calculations are performed using as initial input, the structural parameters of NDCO obtained at 5\,K. In the unit cell, the Nd and Dy atoms are arranged in alternate manner. Thus, each Nd atom has six Dy atoms as nearest neighbours and vice-versa. 
The Nd/Dy $4f$ states were treated as core states for structural relaxation. Ionic positions were relaxed until the forces on the ions are less than 0.1\,meV\AA$^{-1}$. For the electronic self-consistent calculations, the Nd/Dy $4f$ electrons are treated as core and subsequently as valence electrons. The valence band for each atom has the following elements and corresponding orbitals; Cr: $3d$, $4s$, O: $2s$, $2p$ and Nd/Dy: $4f$, $5p$, $5d$, $6s$. The electronic self-consistent calculations are performed in the GGA+U approximation which includes the effect of Coulomb correlation (U). Additionally, the non-collinear calculations are carried out using the GGA+U+SO approximation which includes both Coulomb correlation (U) and spin-orbit (SO) coupling, to find the ground state magnetic order of NDCO system.
\section{\label{sec:Exp}Experimental Results}
\subsection{\label{sec:DC}DC Magnetization}
\begin{figure}[tbh] \center
       \begin{picture}(250,190)
        \put(-5,-5){\includegraphics[width=250pt,height=190pt]{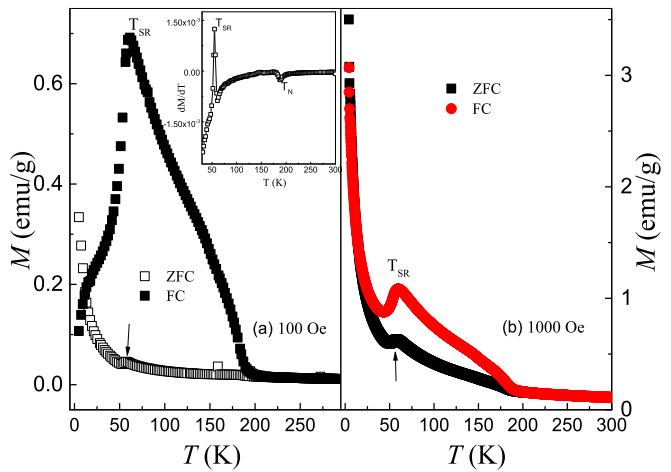}}
      \end{picture}
\caption{(a) ZFC-FC plots of NDCO at 100 Oe, showing bifurcation at $T_\mathrm{{N}}$. Near 50\,K, the arrow marks signature of spin reorientation. The inset shows the derivative (dM/dT) marking $T_\mathrm{{N}}$ and $T_\mathrm{{SR}}$. (b) ZFC-FC plots at 1000 Oe.}
\label{ZFC-FC}
\end{figure}

Fig.~\ref{ZFC-FC}a shows the temperature variation of zero field cooled (ZFC) and field-cooled (FC) magnetization for NDCO from 2\,K to 300\,K in a magnetic field of 100\,Oe. 
The signature of N\'eel temperature ($T_\mathrm{N1}$) occurs just below 200\,K. The $dM/dT$ vs. $T$ plot of the ZFC curve in the inset, shows a dip at 180\,K, corresponding to $T_\mathrm{{N1}}$. With decreasing temperature, the increase in FC curve is rapid, while the increase in ZFC curve is more gradual.
The FC curve shows a continuous increase until it reaches a maximum near 60\,K. In a similar manner, the ZFC curve also shows a peak at 60\,K below which it drops. The simultaneous decrease of ZFC and FC suggests a Morin transition\cite{Prelorendjo1980}. 
However, near 50\,K, the ZFC shows an increase, while the FC falls rapidly. 
The change in ZFC and FC curves can be understood as the process of spin reorientation. The $dM/dT$ vs. $T$ plot also shows the signature of spin reorientation in the inset of Fig.~\ref{ZFC-FC}a.
The transition is similar to that observed in Nd$_{1-x}$Dy$_{x}$CrO$_{3}$ for $x$ $=$ 0.33 and 0.66\cite{McDannald2016}.
Near 20\,K, a crossover between the ZFC and FC curves is observed.

\begin{figure}[tbh]
\begin{picture}(240,190)
\put(-5,-5){\includegraphics[width=240pt,height=190pt]{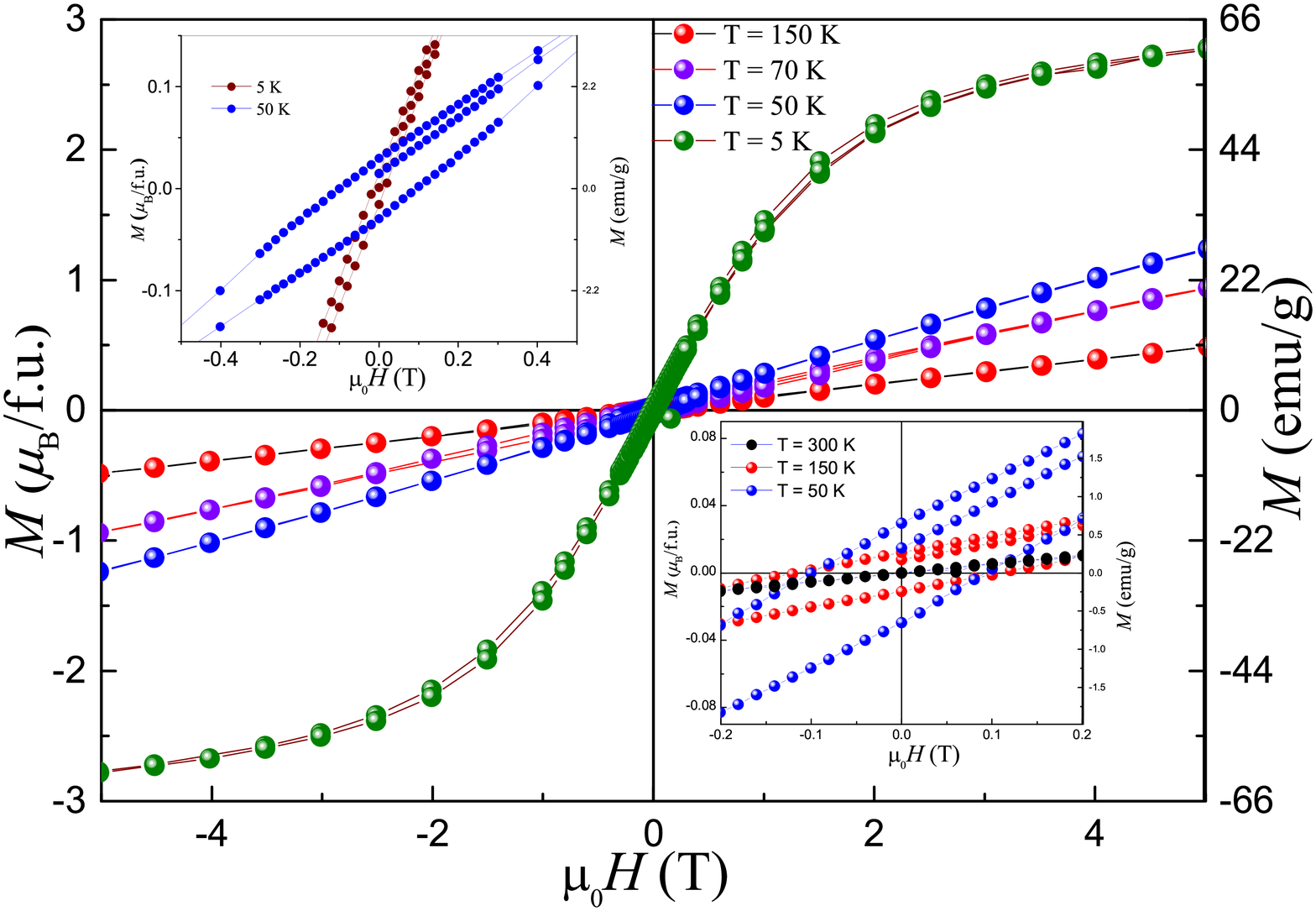}}
 \end{picture}%
    \caption{The magnetization isotherms- M(H) at 150\,K, 70\,K, 50\,K and 5\,K. The lower inset shows paramagnetic behaviour at 300\,K and a weak hysteresis loop at 150\,K just below $T_\mathrm{{N1}}$ and a slightly wider loop at 50\,K. The upper inset shows enlarged low field portion for 50\,K and 5\,K where the coercive field is less than 50\,Oe. }
    \label{MH}
\end{figure}

A crossover, absent in both parent compounds, is seen in Nd$_{1-x}$Dy$_{x}$CrO$_{3}$ for $x$= 0.33 and 0.66\cite{McDannald2016}. In Fig.~\ref{ZFC-FC}b, the ZFC-FC curves are shown in an applied field of 1000\,Oe. A similar bifurcation occurs below 200\,K, although it is of much smaller magnitude. The peak associated with spin reorientation is observed near 60\,K. Below 40\,K, ZFC and FC magnetization values show a continuous rise.
\begin{figure}[tbh]\center
\begin{picture}(240,190)
\put(-5,-5){\includegraphics[width=240pt,height=190pt]{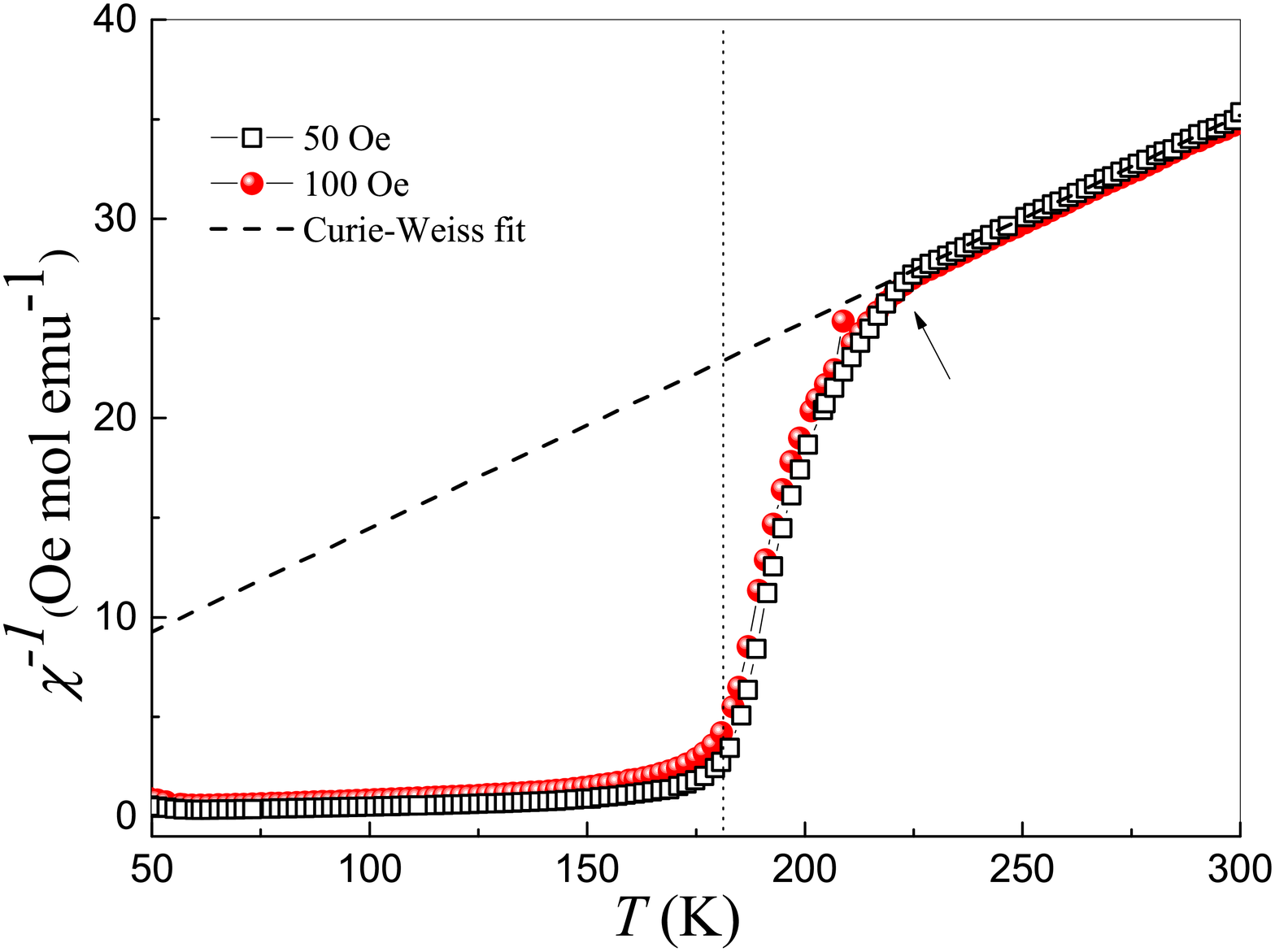}}
\end{picture}%
 \caption{Inverse susceptibility ${\chi}^{-1}$(T) vs $T$ for 50 and 100\,Oe. The dashed line shows the Curie-Weiss fitting for 50\,Oe data with prominent deviation at 220\,K.}
\label{Curie-Weiss}
\end{figure}
\newline 
The temperature variation of inverse magnetic susceptibility, ${\chi}^{-1}$(T) of NDCO in the range 50-300\,K is shown in Fig.~\ref{Curie-Weiss} for fields of 50 and 100\,Oe. The curves show a linear behaviour in the paramagnetic region. However, well above $T_\mathrm{{N1}}$, both the curves show deviation from the Curie-Weiss behaviour. In the presence of 50 and 100\,Oe magnetic field, the deviation starts near 230\,K and is somewhat sharp. 
In the linear region (300\,K to 240\,K), ${\chi}^{-1}$(T)  is fitted to the Curie-Weiss law, ${\chi}^{-1}$(T)=($T$-${\theta}$)/$C$, where $C$=$\mathrm{N_A}$ ${\mu}_\mathrm{{eff}}^2$/3$\mathrm{k_{B}}$, $\mathrm{N_A}$ 
is the Avogadro number, ${\mu}_\mathrm{{eff}}$ is the effective magnetic moment, and $\mathrm{k_{B}}$ is the Boltzmann constant. From our fitting, we obtain ${\theta}$= -20\,K. Though, the negative value conforms to the antiferromagnetic ordering, the small value of ${\theta}$/$T_\mathrm{{N1}}$ ${\sim}$ 0.11, is an indicator of disorder-induced frustration in the system\cite{Chakraborty2017}.
Also, an effective moment (${\mu}_\mathrm{{eff}}$) value of 8.8\,${\mu}_\mathrm{{B}}$ is obtained, which is much higher than the theoretical paramagnetic moment of Cr$^{3+}$ spins (3.9\,${\mu}_\mathrm{{B}}$). However, the total theoretical moment, (0.5${\mu}_\mathrm{{Dy}}^{2}$+0.5${\mu}_\mathrm{{Nd}}^{2}$+${\mu}_\mathrm{{Cr}}^{2})^{1/2}$ = 8.83\,${\mu}_\mathrm{{B}}$ is close to estimated ${\mu}_\mathrm{{eff}}$.
The downward deviation of ${\chi}^{-1}$ is a possible indicator of the formation of Griffith's phase, which is defined as development of FM clusters in a PM matrix and absence of long-range order\cite{Griffiths1969}.
Since the deviation begins near 230\,K it can probably be considered as the Griffith's temperature ($T_\mathrm{{G}}$). In the Griffith's region, ${\chi}^{-1}$ shows a power law behaviour, ${\chi}^{-1}$ = ($T-T_{C}^{R}$)$^{1-{\lambda}}$, where, $T_C^R$ is the maximum possible critical temperature, while ${\lambda}$ is the exponent. In the paramagnetic region, ${\lambda}=$0, while in the Griffith's phase 0 $< {\lambda} {\le}$ 1\cite{Pramanik2010}.
Attempts were made to fit the low field data, as well as that for 1000\,Oe from plots of log$_{10}$(${\chi}^{-1}$) against log$_{10}$($T/T_{C}^{R}-1$). However, ${\lambda}$=0 could not be obtained in the paramagnetic region for H=50\,Oe and 100\,Oe. Although the ZFC-FC suggests formation of clusters above $T_\mathrm{{N1}}$, the clustered phase cannot be called as Griffith's phase. 
\newline
The isothermal field variation of magnetization for various temperatures is shown in Fig.~\ref{MH}. At 300\,K, the $M-H$ curve is linear, indicating a paramagnetic phase (lower inset). 
The $M-H$ curves start showing a non-linear behaviour at low magnetic field values only below 150\,K. The $M-H$ curve shows a hysteresis loop with a coercivity of nearly 1000\,Oe, indicating the existence of weak ferromagnetic moment $F_{x}$ of the Cr$^{3+}$ spins.
Between 50\,K and 5\,K, the magnetization isotherms develop a non-linear behaviour, which can be attributed to the polarization of the $R^{3+}$ moments.
 At 5\,K, the $M-H$ loop is `S'-shaped curve with a negligible loop width. In 50\,kOe magnetic field, the magnetization attains near-saturation and a value of 2.7\,${\mu}_\mathrm{B}/\mathrm{f.u.}$ The value is however smaller than 4\,${\mu}_\mathrm{B}/\mathrm{f.u.}$ obtained in the case of Nd$_{0.5}$Dy$_{0.5}$FeO$_{3}$\cite{Singh2020}.
This also suggests a smaller polarizability of rare-earth ions (Nd$^{3+}$ and Dy$^{3+}$) in the presence of applied field in NDCO.
\subsection{\label{sec:sp.heat}Specific heat}
\begin{figure*}[tbh]
\centering
\includegraphics[width=8.25cm]{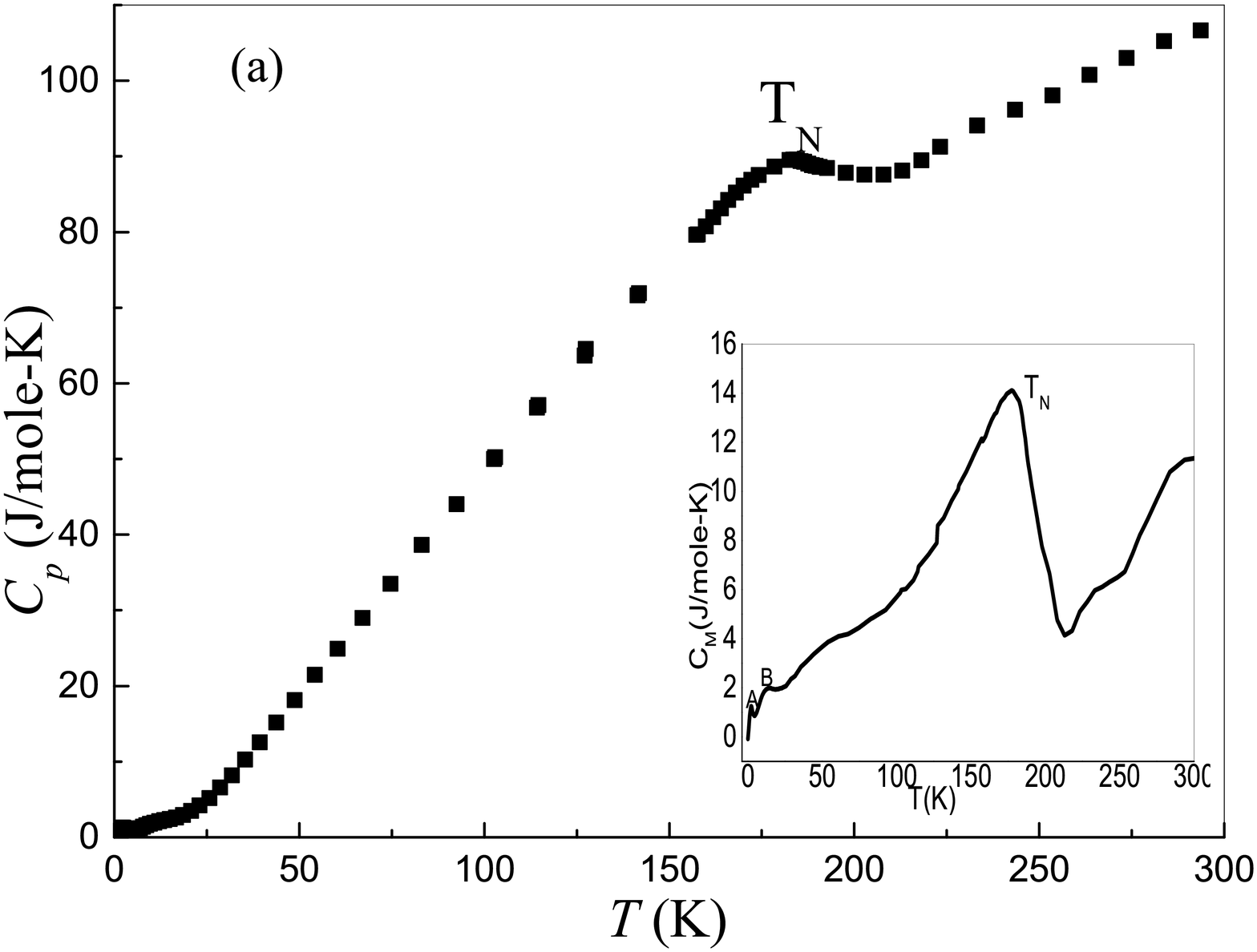}
\hspace*{0.1cm}
\centering
\includegraphics[width=8.25cm]{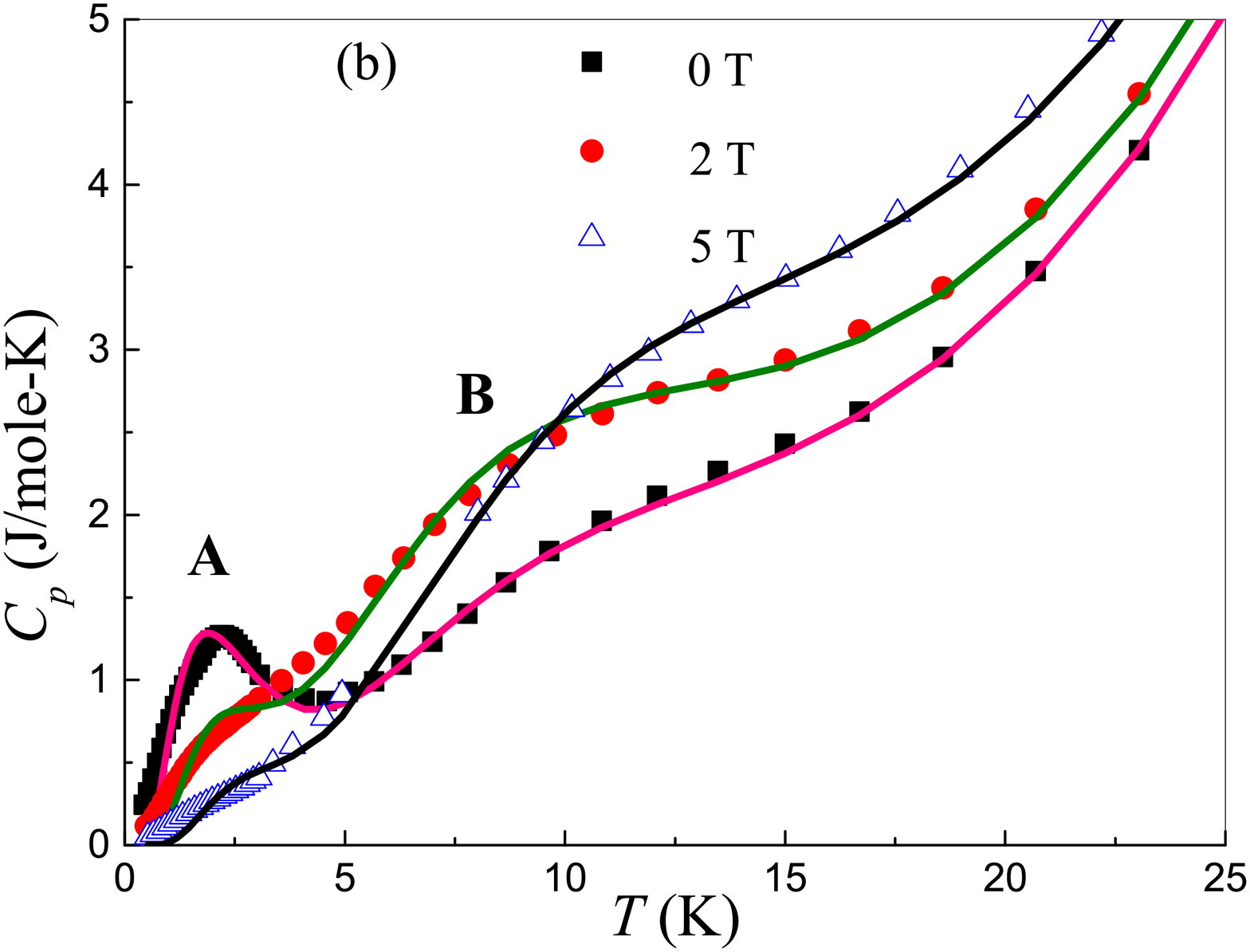}
\caption{(a) Specific heat of NDCO between 0.4\, to 300\,K indicating the N\'eel temperature. The inset shows the magnetic specific heat ($C_\mathrm{{M}}$). (b) Specific heat of NDCO between 0.4\, to 30\,K at zero field and fields of 2\,T and 5\,T. The solid curves indicate fitting to Schottky and lattice specific heat terms.}
\label{specific_heat}
\end{figure*}
  As shown in the Fig.~\ref{specific_heat}a, the specific heat ($C_\mathrm{{p}}$) of NDCO in the temperature range 0.4-300\,K has been measured. To obtain the magnetic specific heat in the entire temperature range, the specific heat of non-magnetic LaGaO$_{3}$ has been subtracted from that of NDCO. The subtraction was carried after a suitable scaling of the specific heat data of LaGaO$_{3}$. The inset shows the magnetic specific heat in the full temperature range. A ${\lambda}$-anomaly in the  magnetic $C_\mathrm{{p}}$ data is observed near 175\,K, which can be considered as the ordering temperature of the Cr$^{3+}$ sub-system, in agreement with the magnetization studies. The width and the height are similar to that reported in NdCrO$_3$\cite{Bartolome2000}. The broader transition observed in our sample can be attributed to the polycrystalline nature of our sample. Additional disorder due to the presence of two rare-earth atoms can also contribute to the broad transition. 
 Below 100\,K, however, we do not observe any signature of spin reorientation in the present heat capacity data, contrary to that observed in NdCrO$_{3}$\cite{Bartolome2000}.

%
In Fig.~\ref{specific_heat}b, we show the low temperature specific heat, in the temperature range 0.4\,K to 25\,K. The gradual broadening of specific heat curve below 20\,K indicates the contribution from the electronic Schottky specific heat due to the crystal field effects in Nd$^{3+}$ and Dy$^{3+}$ ions. 
\newline
A ${\lambda}$-anomaly indicating independent rare-earth ordering is absent.
The low temperature specific heat shows a prominent peak (marked A) at 2.5\,K. Additionally, near 10\,K, a broad hump-like feature marked B occurs.
The features A and B can be attributed to the electronic Schottky specific heat. The feature A signifies the  Schottky specific heat typically seen in various rare-earth based compounds. 
However, the peak height and the width of feature A in NDCO, is smaller than that observed in the similar compound Nd$_{0.5}$Dy$_{0.5}$FeO$_{3}$\cite{Singh2020}. 
Unlike in Nd$_{0.5}$Dy$_{0.5}$FeO$_{3}$, we observe two distinct features A and B, associated with Schottky specific heat\cite{Singh2020}.
The Schottky specific heat can be understood in terms of individual crystal field splitting of both rare-earth ions.
The ground state of Dy$^{3+}$ ion is $^{6}H_{15/2}$ while that of the Nd$^{3+}$ ion is $^{4}I_{9/2}$. Both the ions in monoclinic crystal field $C_{1h}$ split into a series of Kramer's doublets. Thus, the ground state of both ions are Kramer's doublets, which are further split by internal and external magnetic fields. 
Thus, feature A can be attributed to the splitting of the ground state  doublet of the Dy$^{3+}$ ion. The feature B is due to the  splitting in the Nd$^{3+}$ ion, since in NdCrO$_{3}$, a similar feature is observed above 10\,K in specific heat data\cite{Bartolome2000}.
 The specific heat in the range 2 to 25\,K is fitted to the sum of two-level Schottky terms associated with the ground state doublets of Dy$^{3+}$ and Nd$^{3+}$ ions along with the lattice term as given in the equation (1).
\begin{equation}
\label{eqn4}
 C_{p}= \frac{1}{2}R\sum_{i=1}^{2}w_i(\frac{{\Delta}_{i}}{\mathrm{k_B}T})^2\frac{exp\left[-\frac{{\Delta}_{i}}{\mathrm{k_B}T}\right]}{(1+exp[-\frac{{\Delta}_{i}}{\mathrm{k_B}T}])^2}+B_3T^3
\end{equation}
In the above equation, ${\Delta}_{1}$ and ${\Delta}_{2}$ correspond to the splitting in the Dy$^{3+}$ and Nd$^{3+}$ ions, while $B_{3}$ is the lattice contribution.
The fitting is carried out for 0, 2, and 5\,T. At 0\,T, the fitting yields ${\Delta}_{1}$/{$\mathrm {k_{B}}$} = 4.5\,K and ${\Delta}_{2}$/{$\mathrm {k_{B}}$} = 29\,K.
From optical studies on DyFeO$_{3}$, a splitting of nearly 6.5\,K is obtained\cite{Schuchert1969}, which is close to the value obtained in NDCO. 
In a similar manner, ${\Delta}_{2}$/{$\mathrm{ k_{B}}$} is also in agreement with a splitting of 27\,K obtained in the case of NdCrO$_{3}$\cite{Bartolome2000}. 
The clear separation of the two Schottky features is due to the fact that Nd-Cr exchange interactions are much stronger than the Dy-Cr interactions\cite{Bartolome2000,Krynetskii1997}. 
In a field of 2\,T, ${\Delta}_{1}$ increases to 5.9\,K, while ${\Delta}_{2}$ decreases to 26\,K. Finally, at 5\,T, ${\Delta}_{1}$ increases to 8.3\,K, while ${\Delta}_{2}$ also increases to 31\,K. 
The difference between ${\Delta}_{1}$ and ${\Delta}_{2}$ suggests that the Nd-Dy exchange interactions do not play any significant role, which is also evident from our neutron diffraction studies, as discussed below.

\subsection{\label{sec:neutron}Neutron Diffraction}
\begin{figure}\center
\begin{picture}(265,320)
 \put(-5,-5){\includegraphics[width=265pt,height=320pt]{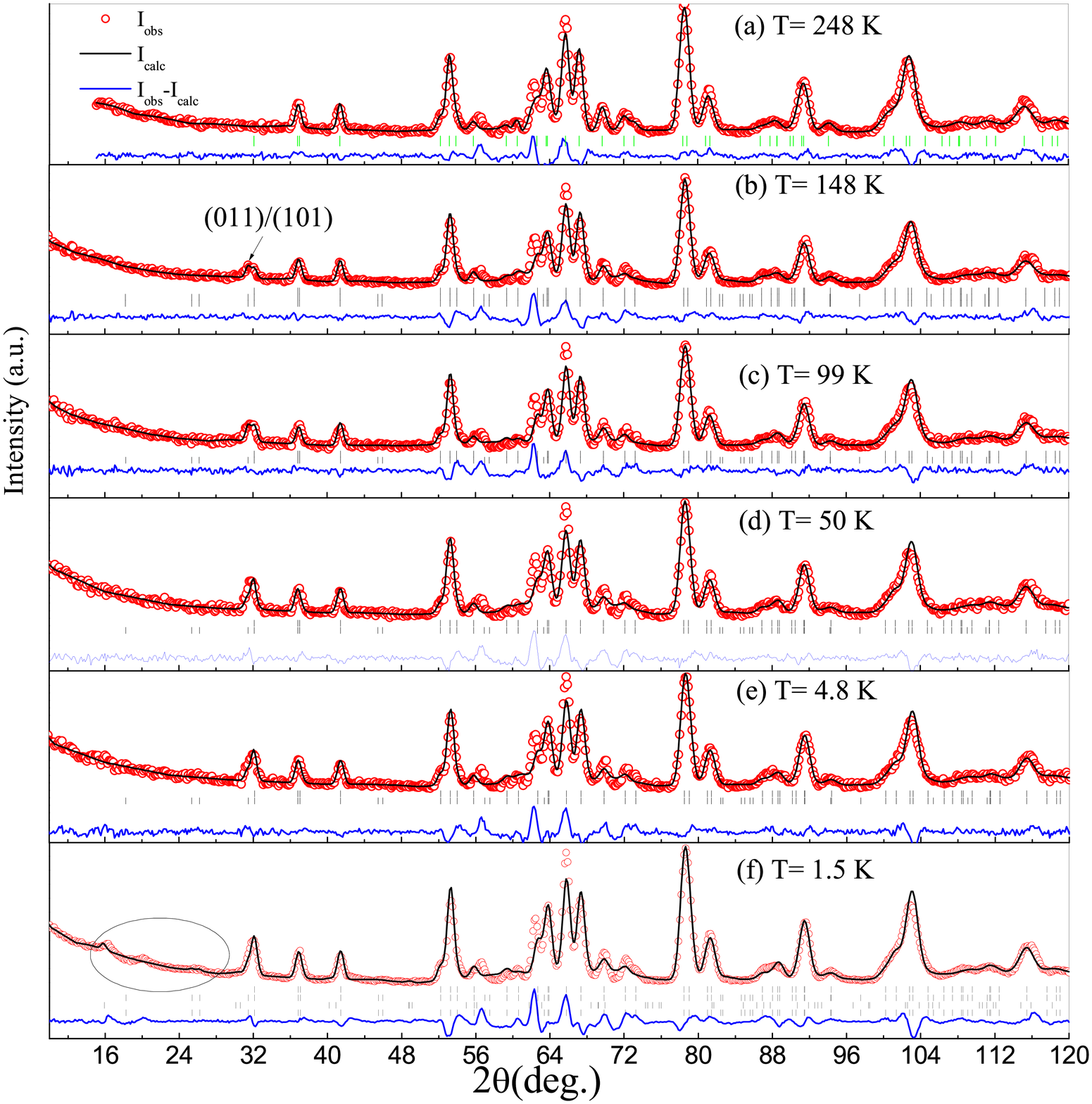}}
\end{picture}
\caption{Powder neutron diffraction pattern and refinements of NDCO at 248\,K, 148\,K, 99\,K, 50\,K, 4.8\,K and 1.5\,K showing the systematic evolution of intensity of (011) and (101) magnetic peaks. Additional peaks due to rare-earth ordering are marked in the loop. }
\label{neutron_diff}
\end{figure}

\begin{figure}[h] \center
       \begin{picture}(240,190)
        \put(-5,-5){\includegraphics[width=240pt,height=190pt]{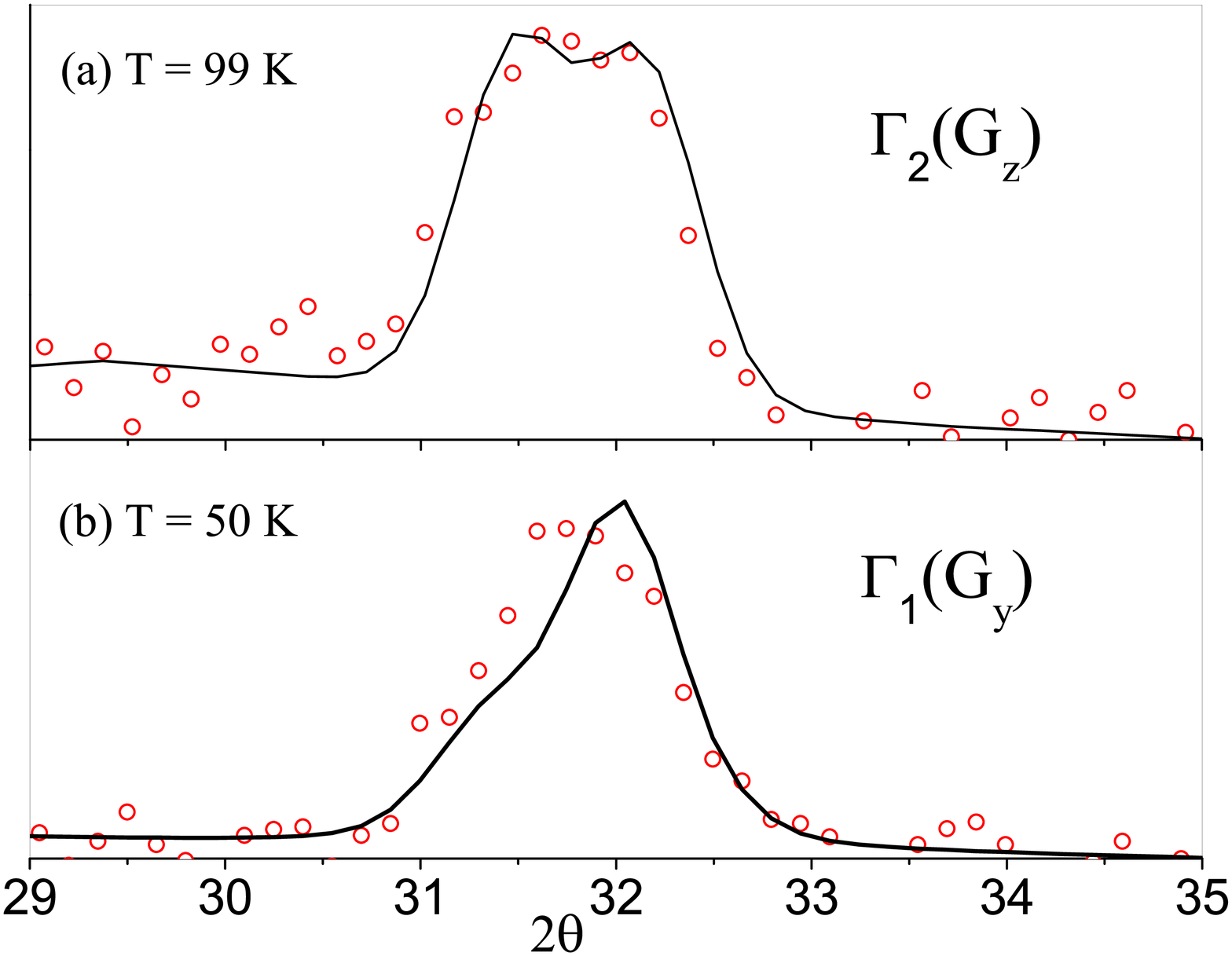}}
       \end{picture}
\caption{Enlarged magnetic Bragg peaks of NDCO for (a) 99\,K (b) 50\,K.}
\label{Neutron_enlarged}
\end{figure}
To ascertain the evolution of magnetic structure and possible rare-earth ordering, powder neutron diffraction experiments were performed in the temperature range 300 - 1.5\,K. 
The diffraction data obtained at 248\,K consists of Bragg peaks which correspond to the crystal lattice (Fig. \ref{neutron_diff}a).
In the diffraction pattern for 200\,K the emergence of magnetic peak around 2${\theta}$=$30^{\circ}$ is observed (not shown), which corresponds to ordering of Cr sub-lattice. The signature of long-range magnetic ordering in NDCO starts at 180\,K, close to $T_\mathrm{{N1}}$.
Fig.~\ref{neutron_diff}b shows the diffraction pattern for T=148\,K. A magnetic peak occurs around $30^{\circ}$, which by the limit of resolution is a convolution of two magnetic peaks. These peaks are indexed as (011) and (101), which are structurally forbidden and correspond to the ordering vector, $\vec{k}$=(0,0,0).
In the $Pbnm$ space group, for $\vec{k}$=(0,0,0), there are eight irreducible representations denoted as ${\Gamma}_{1}$ to ${\Gamma}_{8}$. For the Cr atom which occupies the $4b$ Wyckoff position, four of these representations have zero coefficients.
The four possible representations ${\Gamma}_{1}$ to ${\Gamma}_{4}$ correspond to the Shubnikov magnetic space groups, ${\Gamma}_{1}$ ($Pbnm$), ${\Gamma}_{2}$ ($Pbn'm'$), ${\Gamma}_{3}$ ($Pb'nm'$), and ${\Gamma}_{4}$ ($Pb'n'm$) in Bertraut's notation\cite{bertaut1963magnetism}. In cartesian notation, the four magnetic space-groups can be written in a simplified manner as $A_xG_yC_z$, $F_xC_yG_z$, $C_xF_yA_z$, and $G_xA_yF_z$ respectively. Here, $A$, $C$ and $G$ corresponds to A-type, C-type and G-type antiferromagnetic arrangement of magnetic moments and $F$ denotes the ferromagnetic arrangement. The subscript denotes the direction of magnetic moments, i.e. along the x-, y- or z-axis.
The magnetic structures of both the end compounds are well studied. The Cr$^{3+}$ spins in DyCrO$_{3}$ order below $T_\mathrm{{N1}}$= 146\,K  in ${\Gamma}_{2}$ ($F_{x}$, $C_{y}$, $G_{z}$) structure\cite{Hornreich1978}. The Cr$^{3+}$ spins in NdCrO$_{3}$ order at a higher $T_\mathrm{{N1}}$ value of 220\,K in the  ${\Gamma}_{2}$ ($F_{x}$, $C_{y}$, $G_{z}$) structure\cite{Hornreich1978}.
Based on this, the magnetic peak in NDCO was fitted to  ${\Gamma}_{2}$ structure, which yields a satisfactory fit between 180 and 60\,K. 
 Fig \ref{Neutron_enlarged}a shows the enlarged version of the data at 99\,K corresponding to the $\Gamma_2$ representation.
\newline
 At 58\,K(data not shown), the magnetic structure belongs to the ${\Gamma}_{1}$ ($A_{x}$,$G_{y}$,$C_{z}$) representation. This also coincides with the spin reorientation process that occurs between 75\,K and 60\,K, consistent with our bulk magnetization measurements. 
Fig.~\ref{neutron_diff}d shows the data at 50\,K with the enlarged version of magnetic peak in Fig.~\ref{Neutron_enlarged}b. Based on the shape and refinement of the Bragg peaks, it can be said that the Cr$^{3+}$ spins order in the ${\Gamma}_{1}$ structure below 60\,K. The presence of ${\Gamma}_{2}$ or ${\Gamma}_{4}$ is ruled out since their inclusion yields unsatisfactory results in refinement.
NdCrO$_3$ has been reported to undergo spin-reorientation transition from ${\Gamma}_{2}$ ($F_{x}$, $G_{z}$) to ${\Gamma}_{1}$ ($G_{y}$) around 35\,K\cite{Hornreich1978,Bartolome2000}. Thus, the ${\Gamma}_{2}$ to ${\Gamma}_{1}$ spin reorientation below 60\,K is consistent with that of NdCrO$_{3}$\cite{Hornreich1978}. Therefore, the substitution of Dy$^{3+}$ moments results in the enhancement of the Morin transition temperature as compared to NdCrO$_{3}$. The ${\Gamma}_{1}$ structure of Cr$^{3+}$ spins persist until 1.5\,K.   
This behaviour agrees well with that of Nd$_{0.33}$Dy$_{0.67}$CrO$_{3}$ and Nd$_{0.67}$Dy$_{0.33}$CrO$_{3}$ compositions\cite{McDannald2016}.
\begin{figure}[h] \center
       \begin{picture}(240,190)
        \put(-5,-5){\includegraphics[width=240pt,height=190pt]{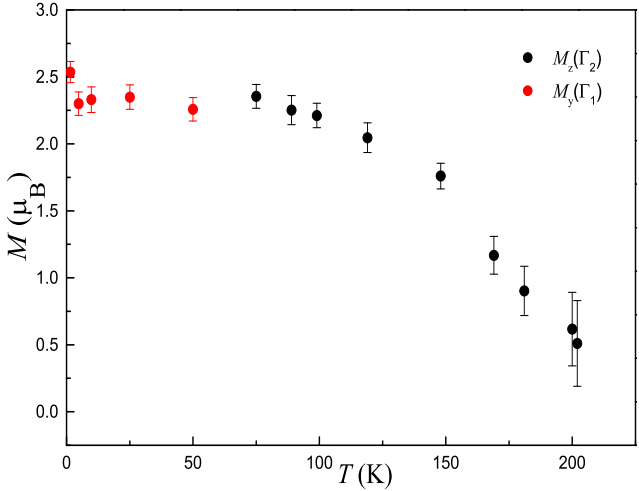}}
      \end{picture}
\caption{Temperature variation of sub-lattice magnetic moment of Cr$^{3+}$ spins in Nd$_{0.5}$Dy$_{0.5}$CrO$_{3}$.}
\label{magnetic_moment}
\end{figure}
\newline
The temperature variation of the magnetic moments for the Cr$^{3+}$ spins for different representations are shown in Fig.~\ref{magnetic_moment}. From 200\,K till 75\,K, we observe an increase in $M_{z}$, the magnetic moment associated with Cr$^{3+}$ spins in the z-direction i.e., $G_{z}$ configuration. Between 75\,K and 50\,K, the $M_{z}$ component of magnetization vanishes due to the spin reorientation. At 50\,K, the magnetic moment is due to $M_{y}$, corresponding to the ${\Gamma}_{1}$($G_{y}$) spin structure. At 1.5\,K, the magnetic moment, $M_y$, is nearly 2.65(5)\,${\mu}_\mathrm{{B}}$, which is slightly lower than the theoretical value of 3\,${\mu}_\mathrm{{B}}$, achieved by complete ordering of the Cr$^{3+}$ spins.
\newline
At 1.5\,K, no long range rare-earth ordering occurs. Since a single peak associated with any predicted long-ranged ordering of Nd/Dy moments is not observed. Also, a single broad hump indicating short-ranged ordering of Nd/Dy moments is also not observed. However, in Fig.~\ref{neutron_diff}f we observe small multiple humps centered at 16$^{\circ}$, 20$^{\circ}$, and 25$^{\circ}$. Due to the weak intensity, a satisfactory refinement could not be carried out. The angular positions of humps at 16$^\circ$ and 25$^\circ$ coincide with Dy-ordering, as in DyCrO$_{3}$\cite{Bertaut1968,VanLaar1971} and $c_z^R$ polarization of Nd$^{3+}$ moments respectively\cite{Shamir1981}.
Thus, the arrangement of the $R^{3+}$ moments either due to a) independent long-ranged ordering or b) polarization due to $R$-Cr interactions is duly suppressed in NDCO unlike in doped orthoferrites. In doped orthoferrite Nd$_{0.5}$Dy$_{0.5}$FeO$_3$\cite{Singh2020}, the low temperature neutron diffraction pattern shows magnetic Bragg peaks associated with $c_{y}^R$, due to rare-earth polarization. 

\subsection{\label{sec:vasp}Density Functional Theory Calculations}
\begin{figure}[tbh] \center
       \begin{picture}(250,300)
        \put(-5,-5){\includegraphics[width=250pt,height=300pt]{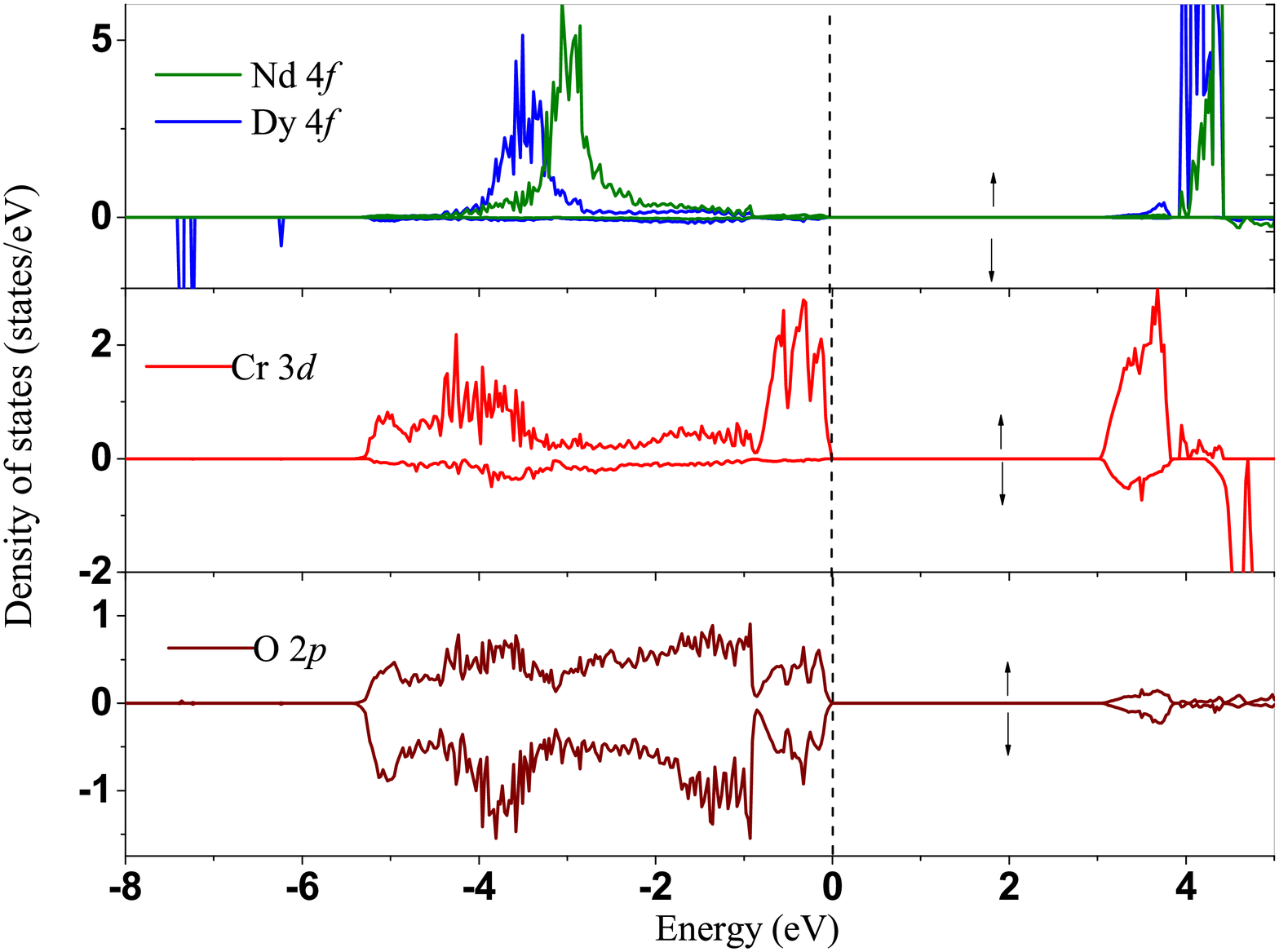}}
      \end{picture}
\caption{Spin polarized partial DOS of NDCO corresponding to $G$-type arrangement of Cr$^{3+}$ and Nd$^{3+}$/Dy$^{3+}$ moments in the GGA+U formalism. The top panel shows partial DOS of Nd/Dy $4f$ states, the middle panel shows the Cr $3d$ states while the bottom panel shows the total O $2p$ states. The Fermi energy is at 0\,eV. The ${\uparrow}$ and ${\downarrow}$ denotes the up and down spin polarized states.}
\label{DOS}
\end{figure}
The electronic structure calculations are performed for various collinear magnetic configurations of Cr sublattice with ferromagnetic (FM), A-type, C-type and G-type anti-ferromagnetic arrangements of the Cr$^{3+}$ moments within the structurally optimized NDCO unit cell. In these calculations, the Nd/Dy $4f$ electrons are treated as core-electrons. Thus there are no 4f moments at Nd/Dy sites. 
For $U=$ 5\,eV and $J=$ 1\,eV, the calculated magnetic moment of Cr$^{3+}$ is found to be 2.9\,${\mu}_\mathrm{{B}}$, which is close to the free ion moment value of 3\,${\mu}_\mathrm{{B}}$.
Comparing the total energies of various magnetic configurations considered, it is found that G-type arrangement has the minimum energy. The total energies have the following sequence $E_{G} < E_{C}$ $<$ $E_{A} < E_{FM}$. Setting the G-type structure as 0\,meV, the energies of C-type, A-type and FM magnetic structures are 30\,meV, 67\,meV and 97\,meV respectively. Thus, we see considerable difference in energy between various magnetic structures of Cr sublattice. Based on the method used by Nagaosa $et. al.$\cite{Nagaosa2004}, the Cr$^{3+}$-Cr$^{3+}$ in-plane magnetic exchange coupling $J_{ab}$ and out-of-plane exchange coupling $J_{c}$ are determined. From our calculations, we obtain $J_{ab}$ $=$ -7.4\,meV and $J_{c}$ $=$ -6.9\,meV, indicating a small anisotropic character. Additionally, our non-collinear calculations within GGA+U+SO approximation with Cr$^{3+}$ spins oriented along different crystallographic directions reveal that for room temperature structure of NDCO at 300\,K, the $\Gamma_2$ configuration(see above) is energetically lower than the $\Gamma_1$ configuration by 0.01\,meV, while for the low temperature (1.5\,K) structure, the $\Gamma_1$ configuration is lower by 0.02\,meV. Thus, from our calculations we observe the spin-reorientation of Cr$^{3+}$ spins even in the absence of $4f-3d$ interactions (though the energy difference is very low) which was found to be essential for another doped orthoferrite Nd$_{0.5}$Dy$_{0.5}$FeO$_{3}$\cite{Singh2020}. 
These results are consistent with our experimental observations as the spin-reorientation of Cr spins is observed while no rare-earth magnetic ordering is observed down to the lowest temperature measured. 
\newline
To probe the possible rare-earth ordering, self-consistent calculations are performed by taking the Nd/Dy $4f$ electrons as valence electrons. The Nd$^{3+}$/Dy$^{3+}$ moments are arranged in G-type, C-type and A-type antiferromagnetic arrangements, while the Cr$^{3+}$ magnetic arrangement is fixed as G-type in all calculations. 
The Hubbard parameters for Nd and Dy are kept at $U$ $=$ 7.5\,eV and $J$ $=$ 0.5\,eV respectively, while for Cr$^{3+}$, the parameters are kept at same values as previous calculations. From the calculations, it is found that the A-type ordering has lowest energy (set as 0 meV) while the G-type and C-type arrangements of $R^{3+}$ moments have energies of +5\,meV and +50\,meV respectively. Thus in NDCO, the C-type arrangement of rare-earth has the highest energy. This is in contrast to the situation in mixed-doped orthoferrite Nd$_{0.5}$Dy$_{0.5}$FeO$_{3}$, where the C-type arrangement of $R^{3+}$ moments is lower in energy as compared to their G-type arrangement\cite{Singh2020}.
Thus, the polarization of Nd$^{3+}$ moments observed in Nd$_{0.5}$Dy$_{0.5}$FeO$_{3}$ due to the Nd-Fe interactions seems to be non-existent in NDCO.
%
\newline
In Fig.~\ref{DOS}, we present the spin-resolved partial density of states (pDOS). The top panel contains the $Nd/Dy$ $4f$ DOS, while the middle and bottom panel display the Cr $3d$ and O $2p$ DOS respectively. 
As seen from the DOS, NDCO is an insulator and has a band gap of nearly 3\,eV within the GGA+U with $U$ and $J$ values mentioned above. An increase in $U$ systematically increases the band gap. 
Below the Fermi energy there occurs significant spectral weight due to Cr $3d$ states, while the O $2p$ has a much smaller spectral weight. The Nd $4f$ states occurs in the region -3 to -4\,eV, while the Dy $4f$ states are split into two distinct regions. The first region corresponds to a broader peak near -4\,eV, while around -8\,eV, the Dy states show extremely sharp and localized peaks. Based on the figure, it can be said that the electronic states between Nd and Dy show minimal overlap.    
%
%
The lack of any overlap between the electronic states of Nd and Dy is in agreement with the neutron diffraction data at 1.5\,K which does not show signature of long ranged ordering, either independently or by polarization due to Cr $3d$ spins.

\subsection{\label{sec:level5}Anomalous Volume Expansion}
The structural studies were carried out in a wide temperature range, from 550\,K to 5\,K in NDCO. In the entire temperature range, the structure of NDCO is analyzed in the $Pbnm$ space group.
\begin{figure}[tbh] \center
\begin{picture}(260,210)
    \put(-5,-5){\includegraphics[width=260pt,height=210pt]{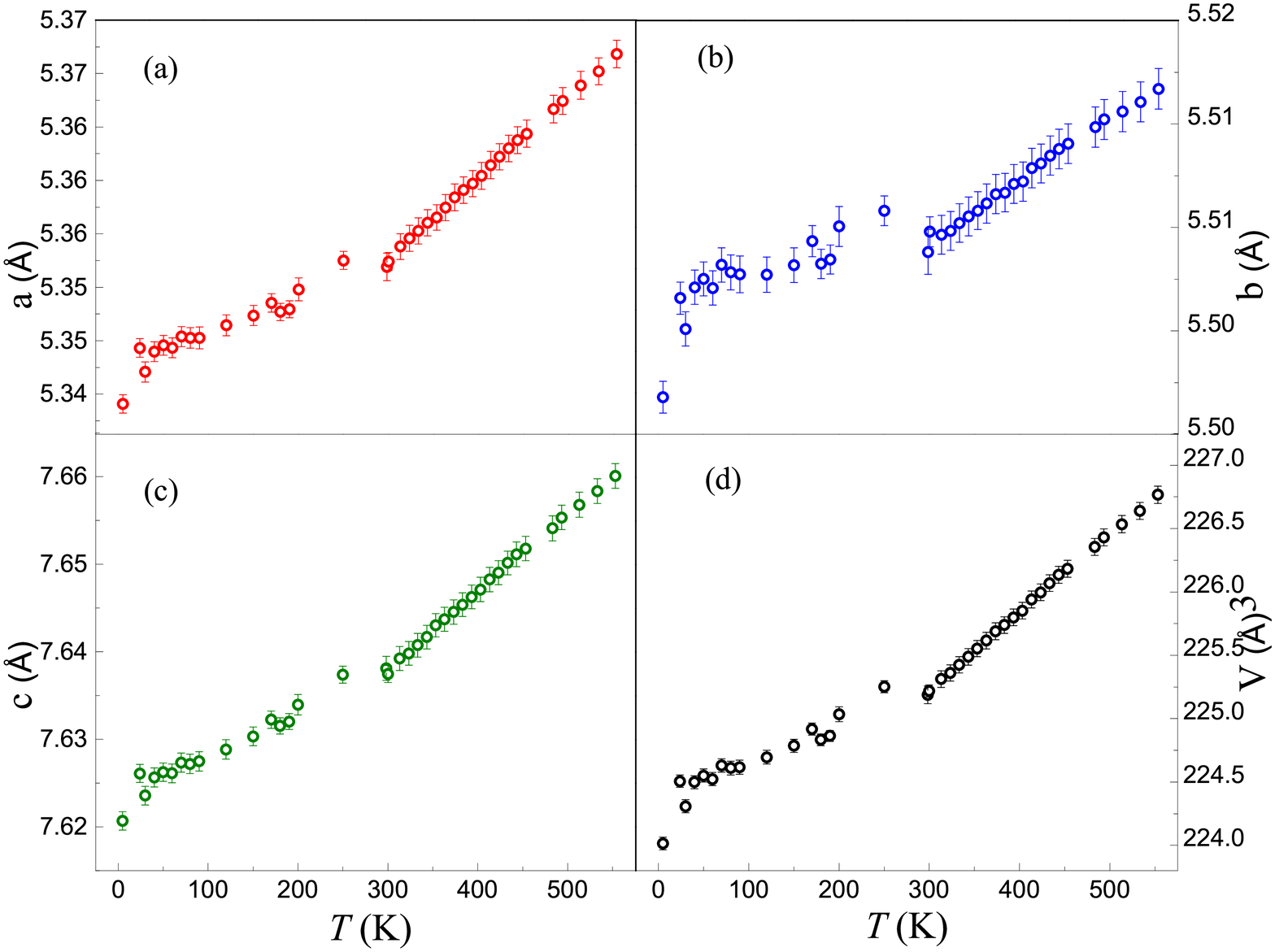}}
\end{picture}
\caption{Temperature variation of lattice parameters and unit cell volume of NDCO.}
\label{Lattice_parameters}
\end{figure}
       
\begin{figure}[tbh] \center
\begin{picture}(240,190)
   \put(-5,-5){\includegraphics[width=240pt,height=190pt]{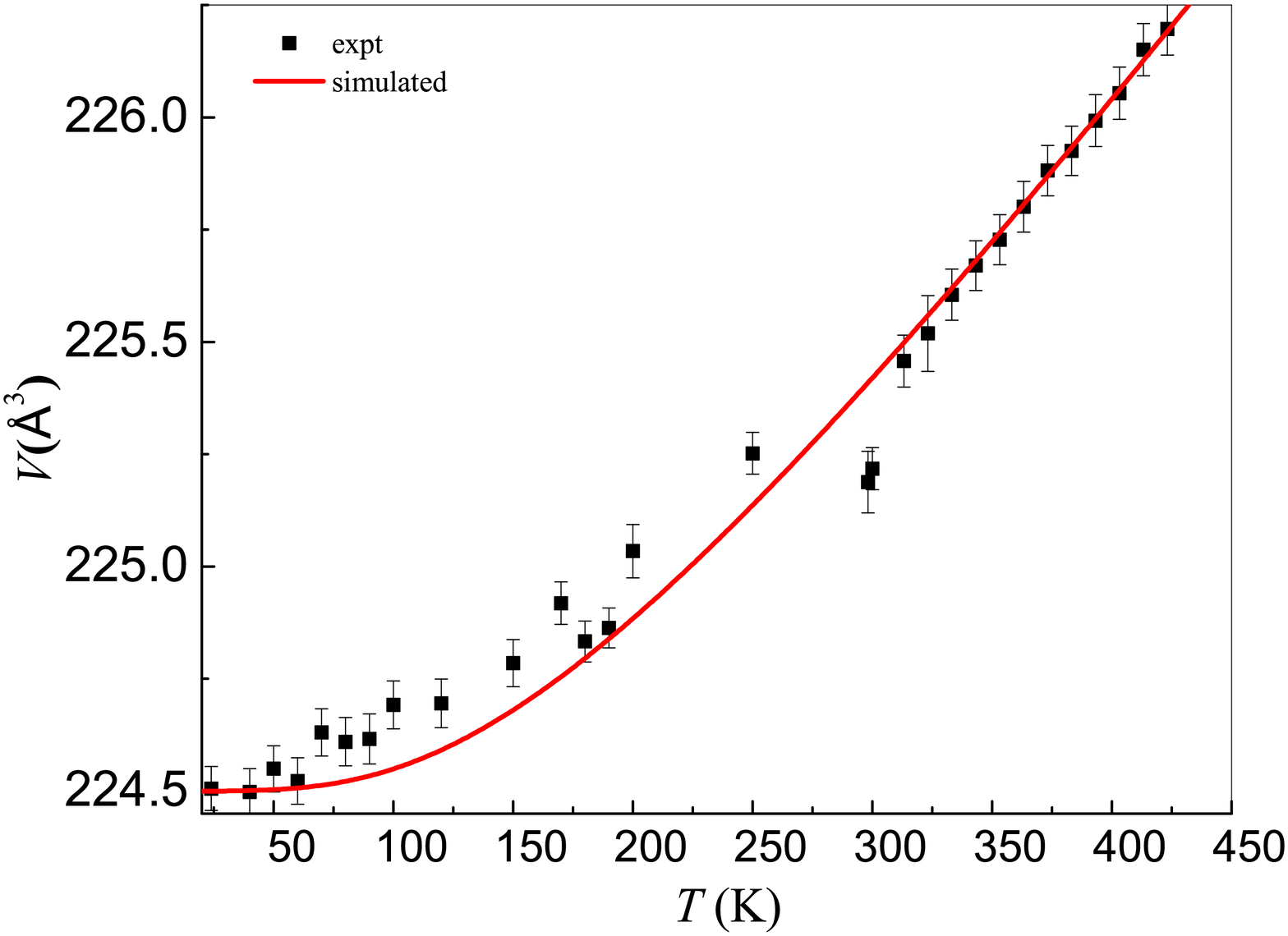}}
\end{picture}
\caption{Temperature variation of the unit cell volume using Gr\"{u}neisen approximation. }
\label{volume_fitting}
\end{figure}

The temperature dependence of lattice parameters obtained from the Reitveld refinement of synchrotron x-ray diffraction data are presented in Fig.~\ref{Lattice_parameters}.
Between 550\,K and 300\,K, all the three parameters show a systematic decrease. Below 300\,K, however, all the three lattice parameters show a step like increment with a peak around 250\,K. The increase is most pronounced for $b$. A corresponding increase is also seen in temperature variation of unit cell volume ($V$), as shown in Fig.~\ref{volume_fitting}. The hump near 250\,K may also be considered as the magneto-volume effect. The negative thermal expansion hints at a polar instability in the system. However, reduction in symmetry is not observed in this temperature region.
In the parent compound DyCrO$_{3}$, a similar increase in volume is observed, with a peak near 260\,K to 280\,K, denoted as $\mathrm{T_{d}}$\cite{Yin2018}. In NDCO, the decrease in volume below 250\,K is more gradual, in contrast to the relatively sharp drop between 550\,K and 300\,K. However, no structural transformation is observed, from our refinement. Attempts to fit the data using $Pna2_{1}$ space group,  yield a higher ``goodness of fit''. 
An increase in the volume is compensated by the off-centric displacement of the Cr$^{3+}$ ions, which results in an increase of covalent energy\cite{Barone2011}.
In general, the displacement of Cr$^{3+}$, which removes the centro-symmetric character of the unit cell, is believed to be responsible for the ferroelectric distortion, which is accompanied by reduction of space-group symmetry of the orthochromites\cite{Yoshii2017}. In our analysis of the diffraction data, attempts to free the position of Cr atom resulted in unusually high displacement from the (0.5,0,0) position.
\newline
In NdCrO$_{3}$, however, the temperature variation of lattice parameters and volume, as well as the ferroelectric distortion occurs well below the N\'eel temperature (at 80\,K), along with lowering of symmetry to $Pna2_{1}$ structure\cite{Indra_2016}. Similar set of transitions are also observed in SmCrO$_{3}$, though well above the N\'eel temperature \cite{Ghosh2014}. This suggests that the nature of the magnetic $R$-Cr interactions play a significant role in the ferroelectric distortion. However, similar symmetry reduction is not reported in DyCrO$_{3}$\cite{Yin2018}.
Thus, in NDCO, we rule out the symmetry lowering associated with a possible ferroelectric transition.
To probe the effects of long range magnetic ordering and magneto-volume effect, the temperature variation of unit cell volume is fitted to the Gr\"{u}neisen equation\cite{Sayetat1998},
\begin{equation}
\label{eqn2}
V(T)= {\gamma}U(T)/K_0+V(0)
\end{equation}

\begin{equation}
\label{eqn3}
U(T)= 9N\mathrm{k_{B}}T(T/{\Theta}_{D})^{3}\int_{0}^{{\Theta}_{D}/T} \left(\frac{x^3}{e^x-1}\right)  dx
\end{equation}
In the above equations, $U(T)$ is the specific internal energy, ${\gamma}$ is the Gr\"{u}neisen parameter, $K_0$ is the incompressibility, $V(0)$ is the extrapolated volume at 0\,K, ${\Theta}_{D}$ is the Debye temperature and $x=\hbar{\omega}/\mathrm{k_{B}}T$. The temperature evolution of unit cell volume along with the theoretical curve is presented in Fig.~\ref{volume_fitting}.
 The theoretical curve deviates from the experimental curve below 300\,K (near 250\,K). Additional deviation occurs below 200\,K coinciding with the magnetic transition. The deviation persists till low temperatures. However, a clear demarcation could not be made between the two regions.
\subsection{\label{sec:level6}Raman Spectroscopy}
\begin{figure}[tbh] \center
       \begin{picture}(240,280)
        \put(-5,-5){\includegraphics[width=240pt,height=280pt]{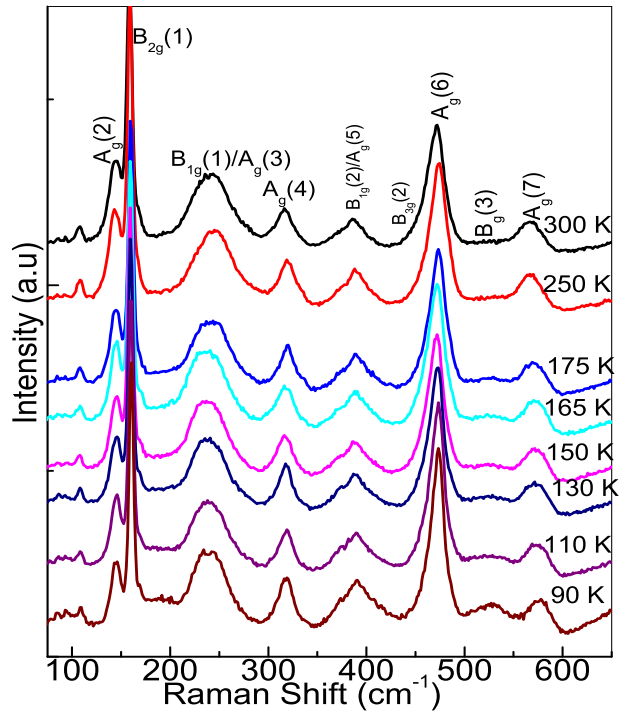}}
       \end{picture}
\caption{Raman spectra of polycrystalline Nd$_{0.5}$Dy$_{0.5}$CrO$_{3}$ for various temperatures between 300\,K to 90\,K. The modes are assigned according to literature. The scales have been shifted vertically.}
\label{RamanSpectra}
\end{figure}
\begin{figure}[tbh] \center
       \begin{picture}(240,260)
        \put(-5,-5){\includegraphics[width=240pt,height=260pt]{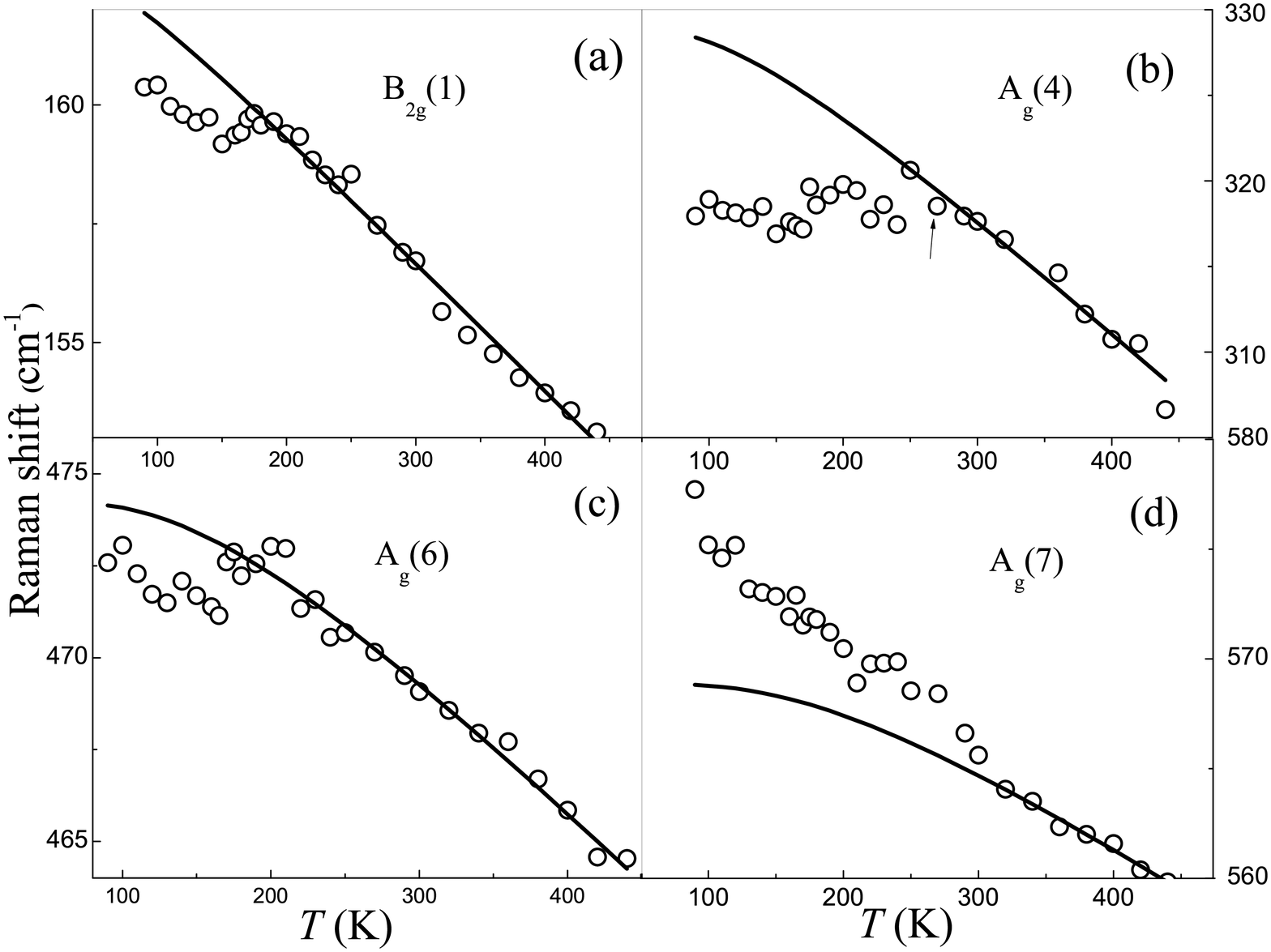}}
       \end{picture}
\caption{Temperature dependence of frequencies of Raman modes (a) $B_{2g}$(1) mode involving pure $Nd/Dy$ vibrations (b) $A_{g}$(4) mode involving Nd/Dy and O vibrations (c) $A_{g}$(6) mode involving the bending of the CrO$_{6}$ octahedra (d) $A_{g}$(7) mode involving the anti-symmetric stretching of the Cr-O bonds. The solid line  represent the fitted curves for anharmonic contributions according to Eq. (4).}
\label{frequencyvar}
\end{figure}

\begin{figure}[tbh] \center
   \begin{picture}(240,190)
   \put(-5,-5){\includegraphics[width=240pt,height=190pt]{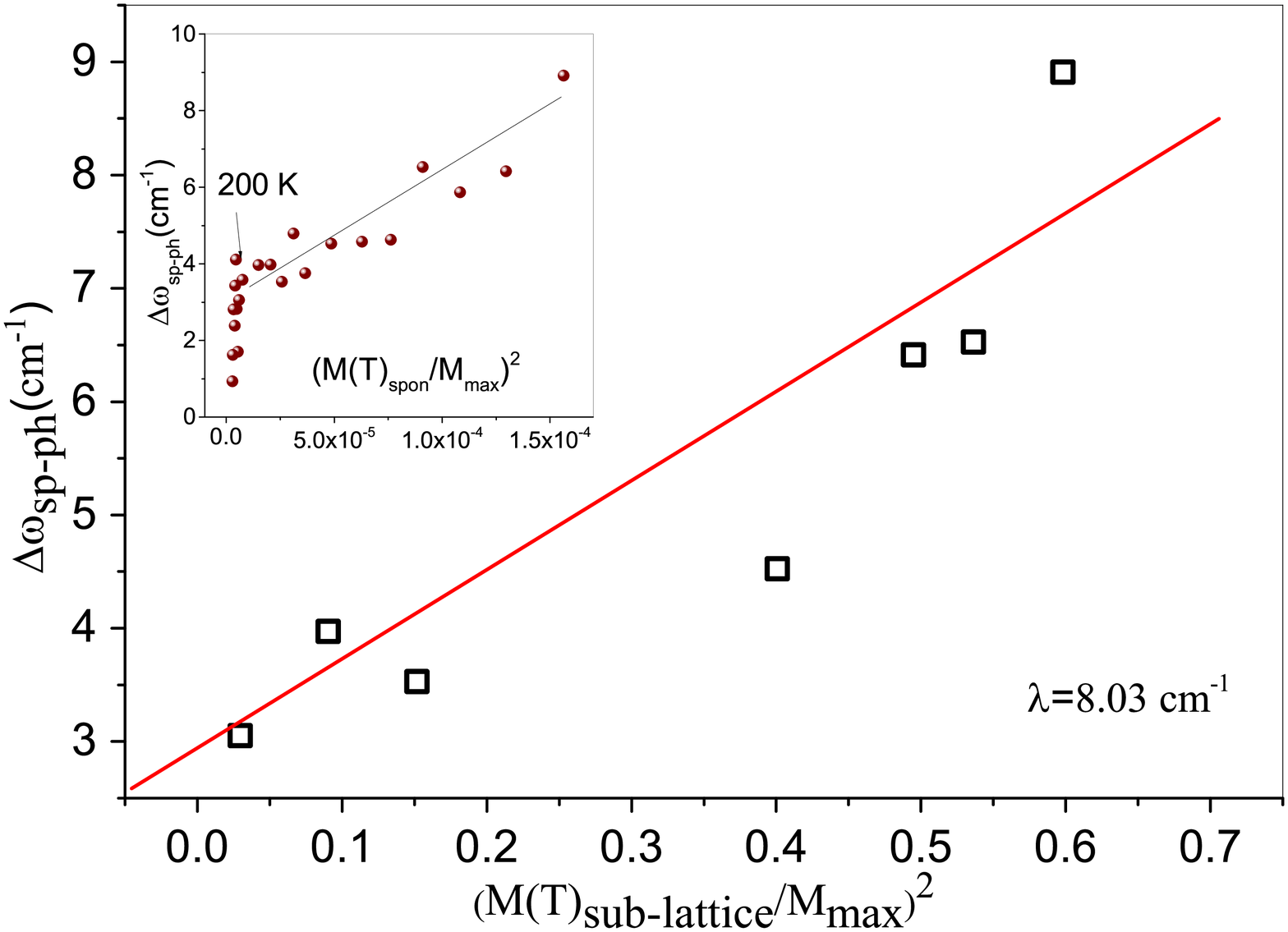}}
   \end{picture}
\caption{Temperature dependent excess frequency ${\Delta}{\omega}$ ($y$-axis) of the stretching mode $A_{g}$(7) between 200\,K and 90\,K. The $x$-axis is the reduced sub-lattice magnetization of the Cr$^{3+}$ spins, obtained from neutron diffraction. The inset shows ${\Delta}{\omega}$ between 90\,K and 250\,K, while the $x$-axis shows the magnetization from the FC curve at 1000\,Oe}
\label{spin_phonon}
\end{figure}
The Raman spectra measured between 450\, and 90\,K is shown in Fig.~\ref{RamanSpectra}. The absence of peak splitting or broadening of the spectra indicates the homogeneous nature of our sample. Unlike previous reports on Dy$_{1-x}$Nd$_{x}$CrO$_{3}$\cite{McDannald2016}, the Raman peaks are sharper with the presence of all the expected peaks. 
The orthorhombic $Pbnm$ structure contains 24 first-order Raman active modes which are classified as  7$A_{g}$ + 5$B_{1g}$ +
7$B_{2g}$ + 5$B_{3g}$, involving vibration of Nd/Dy and Cr, O atoms\cite{SrinuBhadram2013,Iliev1998}. In Fig.~\ref{RamanSpectra}, we show spectra collected both above and below the N\'eel temperature. The modes below 200\,cm$^{-1}$ viz. $A_{g}$(2) and $B_{2g}$(1) can be attributed to vibrations of Nd/Dy atoms. 
The modes above 200\,cm$^{-1}$ can be attributed to the vibrations of the Nd/Dy and the O atoms. 
In the higher energy region of the spectra, the $A_{g}$(7) mode near 560\,cm$^{-1}$ can be attributed to the antisymmetric in-phase stretching vibrations of the Cr-O bonds within the octahedra.
Additionally, the peak near 320\,cm$^{-1}$ corresponding to the $A_{g}$(4) mode is related to the combined Nd/Dy-O vibrations.
It is known that the increase in ionic radii of the $R^{3+}$ ion shifts the modes to lower energy\cite{Weber2012}. Thus, the modes corresponding to Nd-O vibrations would occur at a lower energy compared to Dy-O.
The peak positions are extracted by fitting the individual peaks to the Lorentzian function.
The temperature variation of frequency of various modes is shown in Fig.~\ref{frequencyvar}(a-d). 
To understand the temperature dependence of various modes, the temperature variation of the phonon modes was fitted to the following expression, which corresponds to the anharmonic approximation\cite{Balkanski1983}.
\begin{equation}
\Delta\omega_\mathrm{{anh}} = \omega(0)-C\left(1+\frac{2}{e^{\hbar{\omega}(0)/\mathrm{2k_{B}}T}-1}\right)
\end{equation}

In the above equation, ${\omega}$(0) is the extrapolated frequency (T=0\,K) of the mode, C is the anharmonicity constant. After optimisation, the frequencies of the various modes were fitted in the temperature range 450\,K to 300\,K and extrapolated to 0\,K.
The experimental and theoretical curves are shown in Fig.~\ref{frequencyvar}(a-d). A deviation from anharmonic approximation is seen in all the modes.
In Fig.~\ref{frequencyvar}(a), the temperature variation of $B_{2g}$(1) mode is shown, which corresponds to vibrations of the Nd/Dy atoms. Below 200\,K, there occurs a deviation from the intrinsic anharmonic variation. The nature of deviation indicates softening of the mode, induced by spin-phonon coupling and exchangestriction effect. The latter can result in a net displacement of Nd and Dy atoms. 
\newline
Similar behaviour is also observed in the  temperature variation of $A_{g}$(4) as shown in Fig.~\ref{frequencyvar}(b). The $A_{g}$(4) mode 
corresponds to the Nd/Dy and O vibrations in the $R$O$_{12}$ polyhedra. However, unlike the $B_{2g}$(1) mode, a much stronger anomaly is observed below 300\,K itself. The softening of the mode indicates the influence of the spin-phonon coupling. Similar behaviour is observed in GdCrO$_{3}$, but in case of other rare-earths viz. Sm, Eu, the deviation is relatively smaller\cite{Mahana2017,ElAmrani2014,Bhadram2014}. A combination of the behaviour of $A_{g}$(4) and $B_{2g}$(1) modes suggest a net displacement of the Nd/Dy atoms from their equilibrium positions. However, the different Nd-O and Dy-O bond lengths result in a local variation of the effective displacements and the nature of the softening of the mode.
The $A_{g}$(6) mode also shows softening below 200\,K. However, similar to $B_{2g}$(1), the softening is moderate as shown in Fig.~\ref{frequencyvar}(c). Similar softening is observed in various orthochromites like GdCrO$_{3}$ and SmCrO$_{3}$\cite{Mahana2017,ElAmrani2014}. However, the corresponding deviation is much smaller in both compounds.\newline
In Fig.~\ref{frequencyvar}d, we show the temperature variation of the $A_{g}$(7) mode corresponding to the Cr-O stretching. The solid curve corresponds to the anharmonic contribution to phonon frequency. Interestingly, the deviation of the $A_{g}$(7) mode starts near 300\,K itself, which is nearly 125\,K above the N\'eel temperature. The blue shift indicates hardening of the $A_{g}$(7) mode. 
In GdCrO$_{3}$ and YCrO$_{3}$, similar hardening is observed in the temperature variation of the frequency of $A_{g}$(7) mode\cite{Mahana2017,Sharma2014}. The deviation at a much higher temperature than $T_\mathrm{{N1}}$ indicates the role of magnetovolume effect in NDCO. A similar reduction in unit cell volume has been observed in the case of YCrO$_{3}$\cite{Zhu2020}.
In NDCO, the deviation can be considered as a combined effect of magnetostriction as well as spin-phonon coupling. Based on our magnetization and neutron diffraction studies, the effect of spin-phonon coupling should start to develop only below 200\,K. The spin-phonon coupling in NDCO can be attributed to the R$-$Cr and Cr-Cr exchange interactions. However, in the higher temperature range considered, it is expected that the Cr-Cr exchange interaction plays a greater role in the spin-phonon coupling. Moreover, it is known that the Dy-Cr exchange interactions are negligible.
To probe the effect of spin-phonon coupling, the scaling behaviour of the temperature variation of excess phonon frequency w.r.t. the spin-spin correlation function $<S_{i}.S_{j}>$ is checked. In mean field theory, $<S_{i}.S_{j}>$ is proportional to the square of the sublattice magnetization $(M(T))^2$. Based on the expression derived by Granado et al., the deviation in frequency due to spin-phonon coupling can be written as\cite{Granado1999,Mahana2017}:
\begin{equation}
\Delta{\omega}_\mathrm{{sp-ph}} = \lambda\left(\frac{M(T)}{M(0)}\right)^2
\end{equation}

In the above equation, ${\lambda}$ is the spin-phonon coupling term, $M(T)$ is the temperature-dependent sublattice magnetization from neuron diffraction data, M(0)=3\,${\mu}_\mathrm{{B}}$ is the magnetic moment of Cr$^{3+}$ ion in the totally ordered state at 0\,K.
The spin-phonon coupling constant (${\lambda}$) depends on the mode and the nature of individual atoms, along with the respective bond lengths and bond angles. An estimate of ${\lambda}$ is obtained from Fig. \ref{spin_phonon} which shows a plot of ${\Delta}{\omega}_\mathrm{{sp-ph}}$ versus ($M_\mathrm{{sub-lattice}}$($T$)/M(0))$^2$. In Fig. \ref{spin_phonon}, ${\Delta}{\omega}_\mathrm{{sp-ph}}$ is the excess frequency obtained by subtracting ${\omega}_\mathrm{{anh}}$(T) from the experimental frequencies ${\omega}$(T) of the $A_g$(7) mode.
The value of $M_\mathrm{{sub-lattice}}$($T$) is obtained from total antiferromagnetic sub-lattice magnetization as in Fig. \ref{magnetic_moment}. A nearly linear scaling behaviour is obtained, covering a temperature range of 90\,-200\,K. 
Additionally, to qualitatively check the scaling behaviour, the inset shows the plot of ${\Delta}{\omega}_\mathrm{{sp-ph}}$ and ($M_\mathrm{{spon}}$($T$)/M(0)$)^2$, where $M_\mathrm{{spon}}$($T$) is the spontaneous magnetization, obtained from the FC magnetization for 1000\,Oe.
In this case, the values were considered from 90\, to 250\,K. However, as seen above 200\,K, there occurs no scaling between the excess frequency and magnetization, while below 200\,K, a linear trend is observed. The persistance of the scaling even at 200\,K can be attributed to the magnetic clusters which form well above $T_\mathrm{{N1}}$.
The slope from the main graph corresponds to ${\lambda}$, and has a value of nearly 8\,cm$^{-1}$. The value is considerably higher than the value obtained, for instance in the case of GdCrO$_{3}$\cite{Mahana2017}. A possible reason for the overestimation is an obvious presence of the effects of magneto-volume in the deviation. Additionally, the above formula ignores the R-Cr exchange interactions which can affect the spin-phonon coupling.

\subsection{\label{sec:conc}Conclusion}
Polycrystalline Nd$_{0.5}$Dy$_{0.5}$CrO$_3$ (NDCO) has been synthesised by solid state reaction method. It crystallizes in orthorhombic ($Pbnm$) crystal structure. DC Magnetization shows a  N\'eel temperature of 180\,K, while the large paramagnetic moment and deviation from Curie Weiss behaviour at 230\,K indicates the formation of magnetic clusters.
Powder neutron diffraction measurements indicate that Cr$^{3+}$ moments order as G-type antiferromagnet in $\Gamma_2(F_x,C_y,G_z)$ structure below 180\,K. Below 60\,K, the Cr$^{3+}$ undergoes a spin reorientation transition to $\Gamma_1(A_x,G_y,C_z)$, which persists till low temperature (1.5\,K). Long range rare-earth ordering has not been observed in the powder neutron diffraction data. 
The density functional theory calculations reveal that the G-type arrangement of Cr$^{3+}$ spins is its ground state, while in the case of $R^{3+}$, A-type arrangement is the ground state and C-type magnetic structure has the highest energy.
The absence of hybridization between the Nd and Dy $4f$ states signifies absence of long range ordering of the rare-earth by independent ordering or by polarization due to Cr$^{3+}$ spins.
The heat capacity data shows a $\lambda$-anomaly near N\'eel temperature (180\,K). While the contribution due to rare-earth ordering is absent at low temperature, we observe distinct Schottky anomalies due to Nd$^{3+}$ and Dy$^{3+}$. Low temperature synchrotron XRD measurements rule out the possiblity of a structural phase transition.  However, near 250\,K, the system shows anomalous thermal expansion due to magneto-striction effects. Analysis of Raman spectroscopy data of various modes indicates deviation of the frequencies from the anharmonic approximation at 250\,K itself. The scaling behaviour of the $A_{g}$(7) mode associated with the Cr-O stretching with that of sub-lattice magnetization extends from 90\,K till 200\,K, i.e., even above the N\'eel temperature, which indicates prominent presence of spin-phonon coupling in the system.
\begin{acknowledgments}
MA thanks UGC-DAE-CSR, Mumbai for scholarship (via CRS-M-228). CMNK acknowledges support from the Polish National Agency for Academic Exchange under the `Polish Returns 2019' programme, grant PPN/PPO/2019/1/00014 and the subsidy of the Ministry of Science and Higher Education. RKS is grateful to UGC New Delhi and BHU Varanasi for financial support under IoE scheme. The support of IIT Roorkee through SMILE-13 grant is acknowledged. Characterization facilities of Institute Instrumentation Center, IIT Roorkee are duly acknowledged. 
\end{acknowledgments}
\bibliography{NDCO}

\end{document}